\documentclass[acmsmall]{acmart}

\usepackage{multirow}
\usepackage{subcaption}
\usepackage{float}
\usepackage[whole]{bxcjkjatype}

\def\BibTeX{{\rm B\kern-.05em{\sc i\kern-.025em b}\kern-.08emT\kern-.1667em\lower.7ex\hbox{E}\kern-.125emX}}
    
\copyrightyear{2021}
\setcopyright{acmlicensed}
\acmJournal{PACMHCI}
\acmYear{2021} \acmVolume{5} \acmNumber{CSCW1} \acmArticle{151} \acmMonth{4} \acmPrice{15.00}
\acmDOI{10.1145/3449225}

\newcommand{\secref}[1]{\mbox{Section~\ref{sec:#1}}}
\newcommand{\tabref}[1]{\mbox{Table~\ref{tbl:#1}}}
\newcommand{\figref}[1]{\mbox{Figure~\ref{fig:#1}}}

\newcommand{\nAnabanContentMid}{18,804}
\newcommand{\nAnabanSpeculationMid}{1,253}
\newcommand{\nAnabanContentLate}{97,154}
\newcommand{\nAnabanSpeculationLate}{9,903}
\newcommand{\nTheseusContent}{22,416}
\newcommand{\nTheseusSpeculation}{1,345}

\begin{document}

%
\title[Reaction or Speculation: Building Computational Support for Users in Catching-Up Series]{Reaction or Speculation: Building Computational Support for Users in Catching-Up Series Based on an Emerging Media Consumption Phenomenon}

%
\author{Riku Arakawa}
\orcid{0000-0001-7868-4754}
\affiliation{%
  \institution{The University of Tokyo}
  \city{Tokyo}
  \country{Japan}
}
\email{arakawa-riku428@g.ecc.u-tokyo.ac.jp}
\authornote{These authors contributed equally and are ordered alphabetically.}

\author{Hiromu Yakura}
\orcid{0000-0002-2558-735X}
\affiliation{
    \institution{University of Tsukuba}
    \city{Tsukuba}
    \country{Japan}
}
\email{hiromu@teambox.co.jp}
\authornotemark[1]
\authornote{Also with Teambox Inc., Japan.}

%

%
\begin{abstract}
A growing number of people are using catch-up TV services rather than watching simultaneously with other audience members at the time of broadcast.
However, computational support for such \textit{catching-up users} has not been well explored.
In particular, we are observing an emerging phenomenon in online media consumption experiences in which \textit{speculation} plays a vital role.
As the phenomenon of speculation implicitly assumes simultaneity in media consumption, there is a gap for catching-up users, who cannot directly appreciate the consumption experiences.
This conversely suggests that there is potential for computational support to enhance the consumption experiences of catching-up users.
Accordingly, we conducted a series of studies to pave the way for developing computational support for catching-up users.
First, we conducted semi-structured interviews to understand how people are engaging with speculation during media consumption.
As a result, we discovered the distinctive aspects of speculation-based consumption experiences in contrast to social viewing experiences sharing immediate reactions that have been discussed in previous studies.
We then designed two prototypes for supporting catching-up users based on our quantitative analysis of Twitter data in regard to reaction- and speculation-based media consumption.
Lastly, we evaluated the prototypes in a user experiment and, based on its results, discussed ways to empower catching-up users with computational supports in response to recent transformations in media consumption.
\end{abstract}

%
%


%

%

%
\maketitle

\section{Introduction}
\label{sec:introduction}

As McLuhan \cite{McLuhan:1964} predicted, the Internet has enabled us to communicate across physical distances in ways that have resulted in new forms of media consumption.
One of the first ways that the Internet was leveraged to create new consumption experiences was presented in the context of Social TV \cite{doi:10.1080/10447310701821426}: computer-mediated TV viewing experiences shared by distanced users.
This concept is now commonplace due to the rise of social networking services, especially Twitter \cite{FM3368}.
Moreover, it has been empirically shown that users feel a sense of connectedness to a broader audience in the social viewing experiences by sharing their immediate reactions on Twitter \cite{DBLP:conf/chi/SchirraSB14}.
Vatavu \cite{DBLP:conf/tvx/Vatavu15} took advantage of this finding to enhance users' viewing experiences by proposing an interaction technique to convey the presence of remote users through their silhouettes.

In addition, we are observing an emerging phenomenon in online media consumption experiences in which not just emotional reactions but also \textit{speculation} plays a vital role.
One example of this phenomenon can be found in \textit{Your Turn to Kill} (あなたの番です), a Japanese mystery TV series broadcast by Nippon Television from April to September 2019.
This TV series sparked speculations about what might happen next on social networking services \cite{Nikkan_2019}, where its hashtag repeatedly ranked first in a global ranking of Twitter trends \cite{Cinema_2019}.
Notably, the series' production team intentionally encouraged this movement through marketing efforts, such as holding a promotional event on Twitter using the hashtag ``\#最初の被害者を推理せよ [deduce the first victim]'' during the broadcast of the first episode and posting a summary video about the unsolved mysteries directly prior to the broadcast of the final episode \cite{Toyo_2019}.
In light of events like these, speculation about upcoming episodes of serial media content can be considered to be an important factor in offering better media consumption experiences for users on the Internet. 

These reaction- and speculation-based consumption experiences can both attract users, assuming that they appreciate media content synchronously at the same time that it is delivered.
On the other hand, the Internet has also changed how media content is delivered and consumed.
Specifically, many TV networks now provide catch-up TV services that allows users to watch previous episodes online \cite{Abreu2016}.
This transition suggests that we can no longer assume that media consumption is simultaneous, in contrast to the above discussions for offering better consumption experiences, which rely heavily on synchronous consumption.

We hypothesized that it would be possible to leverage computers to enhance the consumption experiences of such \textit{catching-up users}.
One intuitive approach is to exploit a pseudo-synchronous effect \cite{doi:10.1080/10371397.2013.859982} in the manner Danmaku interaction does \cite{Wu:2019:DNP:3340675.3329485}. 
That is, when a user watches a previously aired episode using a catch-up TV service, tweets posted during the initial broadcast are presented in relation to playback time within the episode.
Considering the effect of Danmaku \cite{DBLP:journals/chb/LimHKB15} and Twitter-based social viewing experience \cite{DBLP:conf/chi/SchirraSB14}, it would make users feel as if they were watching simultaneously with other users and enhance their sense of social presence.
However, we argue that this approach does not sufficiently incorporate media consumption experiences involving speculation, as we mentioned above.
This is because, besides such social aspects, online speculative discussions may have informative aspects, which have not been focused on by this approach.
Moreover, online speculative discussions often contain spoilers \cite{gray2007speculation}, which should be taken into consideration when developing computational supports for catching-up users.

With these points in mind, we investigated the current status of online media consumption experiences, especially for serial media content; evaluated possible approaches to providing better experiences for catching-up users; and discussed possibilities for future computational support.
The steps of our research were as follows:
\begin{enumerate}
    \item We first conducted semi-structured interviews to identify how users involve with and are affected by online media consumption experiences centering on speculation.
    \item We then performed quantitative data analysis of tweets about two TV series to provide background for developing computational supports for catching-up users, illustrating the unique aspects of reaction- and speculation-based media consumption.
    \item Based on the results of (1) and (2), we carried out a user experiment and evaluated the effects of two different approaches to enhancing the consumption experiences of catching-up users.
\end{enumerate}
Our findings and subsequent discussions illustrate further possibilities for computationally supporting catching-up users, especially with regard to speculation, which has received little research attention to date despite its effectiveness in enhancing users' consumption experiences.

\section{Background}
\label{sec:background}

To situate our work, we begin by examining the previous literature on catch-up TV, which has highlighted the increase in the number of catching-up users.
We also review the literature on social experiences in the context of media consumption and discuss the need for interaction techniques to support catching-up users.

\subsection{Increase in Catching-Up Users}
\label{sec:increase-catching-up-users}

TV from the so-called ``network era'' (1952 to the mid-1980s) was long considered a one-way communication medium with limited programming choices, where viewers had to base their daily duties around the schedule that a few TV networks had mandated \cite{Lotz2009What}.
This relationship between TV programs and the audience was drastically altered by the advent of digital video recorders.
Studies were conducted in many countries \cite{Kalia2014,Medina2015} to investigate the impact of digital video recorders, which ultimately found their increased usage \cite{Wilbur2008}.
Following the success of digital video recorders, technologies for enabling other viewing practices---such as downloading, streaming, and mobile viewing---were developed and became widespread \cite{Bury2013}.

In recent years, it is becoming difficult to assume that many people enjoy media content at the time that it is distributed.
Instead, a growing number of users are choosing to catch up after the first air using a catch-up TV website hosted by the networks or on other video-distribution services such as Amazon, Netflix, and Hulu \cite{Matrix_2014, doi:10.1177/0163443717736118}.
According to Vanettenhoven and Geerts \cite{Vanattenhoven2015}, users of video-on-demand services prefer movies and drama series to news and comedies.
The authors conducted in-home interviews to investigate the use of on-demand platforms and reported users' preferences.
They also suggested that users might appreciate replays of episodes in a previous season before the beginning of a new season.
Considering the demand for catching up on the contents of series, we suspect that there is an excellent opportunity to address the needs of catching-up users through computer-supported approaches.

\subsection{Social Experience in Media Consumption}
\label{sec:internet-based-social-interactions}

On the other hand, we argue that the existing computer-supported approaches which highlight social experiences are insufficient to address the current experiences of media consumption in the light of the increase in catching-up users.
To situate our research and develop ideas for improvement, we review the articles about social experiences in media consumption from two aspects: sharing immediate reactions and sharing speculations.

\subsubsection{Sharing Immediate Reactions}
\label{sec:sharing-immediate-reactions}

One popular approach to exploiting computers for media consumption is invoking the feeling of watching TV with peers by sharing reactions in real time.
For instance, Ducheneaut et al. \cite{doi:10.1080/10447310701821426} investigated social interactions among TV viewers and developed the concept of ``Social TV'' that would enable geographically distributed viewers to communicate with one another.
Though their experimental design involved only audio relays between two rooms or pre-recorded videos of previous viewers to simulate the idea, many Internet-based interaction techniques have since been proposed for social TV \cite{Cesar2009,DBLP:conf/ccnc/CesarG11}.
As Cesar and Geerts \cite{cesar2011understanding} listed, this approach has also been implemented in commercial products and services.
To further enhance viewer experiences in this manner, Vatavu et al. \cite{DBLP:conf/tvx/Vatavu15} introduced audience silhouettes for TV by overlaying viewers' body movements on the content in real time.

In addition, the advent of social networking services has promoted social viewing experiences on the Internet \cite{FM3368,Christopher2014}.
Kim et al. \cite{DBLP:conf/chi/KimKKKKO15} analyzed the dynamics of fans on Twitter during the 2014 FIFA World Cup and suggested that fans' social communication on Twitter induced feelings of social presence on a global scale.
Schirra et al. \cite{DBLP:conf/chi/SchirraSB14} reported the existence of Twitter users who actively posted during TV series broadcasts and conducted in-depth interviews with them to understand their motivations for posting.
They concluded that these so-called ``live-tweeters'' seek social viewing experiences that evoke feelings of being connected with others.
These results are also supported by Kim et al. \cite{DBLP:journals/tele/KimMY19}, who conducted an online survey which found that social presence plays a mediator role in the process whereby social viewing experiences lead to viewer enjoyment.
These interaction techniques, however, are not directly applicable to enhancing the consumption experiences of catching-up users, as we can no longer assume that peers are watching the same content simultaneously.

\subsubsection{Sharing Speculations}
\label{sec:sharing-speculations}

Although speculation during media consumption has recently become popular, as we mentioned in \secref{introduction}, little research has focused on it to date \cite{Jenkins2006,gray2007speculation,Mittell2009}.
Jenkins \cite{Jenkins2006} first reported the existence of an online fan community in which speculations were actively posted in the context of the narrative hook of the TV show \textit{Survivor}.
Based on his account, Gray and Mittell \cite{gray2007speculation} conducted an in-depth investigation of speculative discussions on a fan forum of \textit{Lost}, which they described as the most elaborated and narratively complex TV program of the 2000s.
Their online qualitative survey designed to understand fans' motivations for participating in such discussions offered several explanations, including that fans regard the speculations themselves as enjoyable texts.
Mittell \cite{Mittell2009} reported that the \textit{Lost} fan wiki had served as a place to store such speculation.

However, the existing literature has not explored whether catching-up users can appreciate such stored discussions or how to design interactions that leverage speculation for them.
In addition, it is possible that social networking services, which emerged after the aforementioned studies, have significantly transformed the ways of media consumption related to speculation.
Accordingly, an up-to-date investigation is warranted.

Here, we note that our focus is inconsistent with those of previous studies regarding user comments and reviews in media consumption \cite{dellarocas2006motivates,DBLP:conf/cscw/HemphillO12}.
As Gray and Mittell \cite{gray2007speculation} stated, speculation is not a retrospective experience (for example, posting a review on the Internet after watching a movie); rather, it is an ongoing experience based on a form of seriality that is predicated on anticipating the next episode.

\subsection{Interaction Techniques for Supporting Catching-Up Users}
\label{sec:interaction-techniques}

Although a lot of previous studies have explored computer-supported media consumption, few have been proposed and designed explicitly for catching-up users.
One such interaction technique is Danmaku \cite{Wu:2019:DNP:3340675.3329485}, a new commenting system popularized in Asian countries that offers social interactions to viewers of online videos by leveraging pseudo-synchronicity.
In Danmaku, comments are recorded in connection to a specific playback time within a video, indicating when the comments are typed.
The comments are then overlaid on the video in synchronization with future playbacks.
While this is not exactly a synchronous co-viewing experience, Danmaku commenting is known to create a pseudo-synchronized viewing experience \cite{doi:10.1080/10371397.2013.859982}, which leads to a sense of being socially connected \cite{DBLP:journals/ijhci/ChenGR17}.

However, Danmaku was not originally designed for catch-up TV, and there is room to discuss how Danmaku-like interactions can contribute to offering better consumption experiences to catching-up users.
In particular, considering the positive effect of the Twitter-based social viewing experiences \cite{DBLP:conf/chi/SchirraSB14,DBLP:journals/tele/KimMY19}, we can suppose that presenting live-tweets in a Danmaku interface would enhance their consumption experiences, which, however, is not experimentally confirmed.
Moreover, considering that Danmaku is mainly used to share immediate reactions to scenes in videos \cite{DBLP:journals/ijhci/ChenGR17}, we anticipate that simply applying a Danmaku interface to sharing comments on videos would not address the unique particularities of sharing speculations.

Another possible solution is a companion application to help viewers understand complex narratives, such as that proposed by Silva et al. \cite{DBLP:conf/tvx/SilvaATSRM15}.
Similarly, Nandakumar and Murray \cite{DBLP:conf/tvx/NandakumarM14} confirmed that their second-screen application displaying a story map improved the first-time viewers' comprehension of the latest episode of a TV series that they had never watched.
Yet while this technique can effectively summarize information contained in missed episodes without the need to watch those episodes, it is not directly applicable to catching-up users who want to enjoy watching the missed episodes from the beginning.

Therefore, in this paper, we aim to develop computational support for catching-up users by focusing on the effects of sharing immediate reactions and speculations.
We discuss the characteristics of each through semi-structured interviews and analyzing public data and propose new approaches, whose effectiveness is evaluated via a user experiment.

\section{Semi-Structured Interviews on the Effects of Engaging with Speculation}
\label{sec:interview}

Our literature review revealed that previous studies have paid little attention to media consumption related to speculation, especially with reference to interaction techniques.
On the other hand, as mentioned in \secref{introduction}, we are observing that speculation plays a vital role in recent media consumption and anticipating that providing opportunities for catching-up users to relate to speculation would enhance their consumption experiences.
For this purpose, we first conducted semi-structured interviews to understand how people are engaging with such speculation on the Internet during media consumption experiences.
If they are not appreciating speculation, providing opportunities for engaging with speculation would not be beneficial for catching-up users.
In addition, if they are not having trouble relating to speculation in catch-up situations, developing a new computational approach would not be necessary.
In this section, we describe our procedure and findings of our interviews.

\subsection{Participants}
\label{sec:interview-participants}

Our interviews included 10 participants without compensation aged between 23 to 34, of whom three were female.
The nationalities of all participants consisted of East Asian countries (Japanese, Korean, and Chinese).
They were recruited via word of mouth and online communication in a local community where over 100 university students gather.
At the time of the recruitment, all participants self-reported they had experience in engaging with online speculative discussions during media consumption.

\subsection{Methodology}

The interviews were conducted face-to-face, except for four participants who were interviewed via video call.
The interviews were conducted in Japanese, as all participants were fluent Japanese speakers, and were audio-recorded.
Each interview lasted approximately 30 minutes.
Interview questions were designed to explore participants' behaviors in online speculative discussions, including motivation, timing, and feelings and thoughts.
For example, we asked, ``How do you engage with speculation, e.g., reading or posting?''; ``How often do you engage with speculation?''; ``What is your motivation or aim of engaging with speculation?''; ``What is on your mind when you engage with speculation?''; ``How do you feel when you are engaging with speculation?''; ``What do you care about when you engage with speculation?''; and ``What is your usual feeling after engaging with speculation?''

Our analysis was guided by previous research on media consumption-related social interactions on the Internet \cite{DBLP:conf/chi/SchirraSB14,DBLP:conf/ACMdis/HillmanPN14}.
Using open coding \cite{StrCor90}, the transcriptions of the interviews were analyzed to document the behaviors and opinions of participants with regard to speculation on the Internet.
First, one author read through the transcriptions carefully and extracted emergent themes.
All authors then reviewed and organized the themes found in the interviews and discussed apparent discrepancies until a consensus was reached.
Through this iterative refinement process, three key themes were identified, which are outlined below.

\subsection{Findings}
\label{sec:interview-findings}

Our analysis of the interviews revealed several aspects in regards to how the participants are affected by engaging with speculation on the Internet.
In this section, we describe each aspect and present relevant quotes from them.

\subsubsection{Deepening Understandings of Media Content}
\label{sec:interview-understanding}

In our interviews, most participants mentioned that deepening their understandings of media content is one of the motivations or values of engaging with online speculative discussions:
\begin{quote}
    When I want to know how other people interpret the storyline or scenes, I use the Internet to look for speculation. As they remind me of the creators' intentions and hooks that I didn't notice, the value of the content increases much more. (P1)
\end{quote}
\begin{quote}
    For media content which has connotations behind it, I am sometimes confused about how to interpret their descriptions. In such cases, I use Google to look for fans' speculation on them. (P2)
\end{quote}
\begin{quote}
    It is difficult to be confident that I'm enjoying 100\% of the creator's messages by just reading or watching the content once. So, I wonder if there is a way to enjoy [the content] that I haven't noticed, and then, I go looking for speculation. I google the title with the word ``考察'' [meaning ``speculation'' in Japanese] and jump to them. (P10)
\end{quote}
These comments suggest that users' sense that they lacked understanding of media content led them to actively engaging with speculation---for example, searching with the title on the Internet.
They then deepened their understandings or interpretations of the story or connotations of the content, which in turn yielded an enhanced consumption experience.

In addition, participants mentioned various platforms they used to deepen their understandings of media content, such as:
\begin{quote}
    I sometimes refer to personal blogs on the Internet on which long articles involving deep speculation about various content are posted. (P4)
\end{quote}
\begin{quote}
    I like reading long speculation posts on Twitter sharing screenshots of note-taking apps. I have a private list of Twitter users whom I trust in the quality of their speculations and often look for their comments when I finish the latest episode. (P7)
\end{quote}
 \begin{quote}
     I usually visit Twitter to search with hashtags for other users' thoughts. But, in some cases, such as when I couldn't find it interesting though its reputation was good or when I got confused in its interpretations, I googled the title and read some speculating articles. (P10)
 \end{quote}
\begin{quote}
    I often use Instagram to access the speculations of other users. Searching the title of the TV series provides me a stream of many posts. I think the photo accompanying each post reflects the sense of its author well. (P8)
\end{quote}
These comments imply that speculative discussions available in online platforms can be leveraged for constructing computational support for catching-up users.

\subsubsection{Feeling a Sense of Connectedness with Others}
\label{sec:interview-connectedness}

At the same time, we observed that participants felt a sense of connectedness with other users through engaging with online speculative discussions, in the same manner as live-tweets \cite{DBLP:conf/chi/SchirraSB14}:
\begin{quote}
    Seeing reactions to my post of a link to speculation articles or joining online discussions lets me feel an atmosphere of excitement. (P3)
\end{quote}
\begin{quote}
    It is simply fun to discuss online with people who watched the same content. Sometimes I'm convinced by seeing different opinions from mine, but rather, I feel happy when I confirm that someone shares the same feelings or interpretations. (P9)
\end{quote}
\begin{quote}
    Just after watching videos, I often search the titles on Twitter. It's not only about pursuing a novel insight or interpretation. But just diving into other users' thoughts or speculations makes me bask in the afterglow. (P4)
\end{quote}
\begin{quote}
    On Twitter, I can see speculations posted just after the broadcast. I like this sense of liveness with other users. (P7)
\end{quote}
These comments revealed an effect that resembled observations about \textit{Lost} fan forums \cite{gray2007speculation}: These speculative discussions not only provide opportunities to reread content with a more in-depth analysis and enhance narrative pleasure, but also offer a communal relationship that circulates such discussions.

In addition, we found that user behaviors could be divided into ``posters'' and ``lurkers,'' as previous studies of online communities \cite{DBLP:journals/chb/SchneiderKJ13,10.1145/3272973.3274089} have suggested.
That is, posters are actively involved in speculation and discussion with other users, such as the first two quotes from P3 and P9 above indicate, whereas lurkers merely consume speculations posted by other users, as indicated by the last two quotes from P4 and P7.
Lurkers noted that they also felt a sense of connectedness with others through reading speculative discussions.

In summary, our interviews demonstrated commonalities with the behavioral types and effects described in previous studies of social experiences in media consumption.
In addition, we elucidated an unrevealed aspect that would not achieved by sharing immediate reactions, that is, deepening understanding through online speculative discussions.
These positive aspects delivered by engaging with online speculative discussions suggest that providing opportunities for catching-up users to relate to speculation would enhance their consumption experience.
On the other hand, our interviews also revealed a side effect regarding this phenomenon, which in turn require consideration when constructing computational support.

\subsubsection{Concerns About Encountering Spoilers}
\label{sec:interview-spoilers}

Participants also mentioned a certain side effect of engaging with online speculative discussions:
\begin{quote}
    While I'm not caught up with the latest episode, even though I didn't search the title, someone on my timeline tweeted about the speculation. Such a spoiler causes me to lose motivation to continue watching. (P4)
\end{quote}
\begin{quote}
    I felt quite a bit of regret when I saw spoilers for content that has an elaborate storyline. (P5)
\end{quote}
\begin{quote}
    Sometimes when I searched the names of characters, I happened to see images containing spoiling speculation. (P2)
\end{quote}
According to Gray and Mittell \cite{gray2007speculation}, some users prefer to be spoiled because spoilers enable them to take control of their emotional responses.
However, the participants in this study expressed complaints about spoilers or at least wished to control their exposure to such information.

In particular, some participants mentioned their hesitation to see speculations due to the risk of being spoiled when they were catching up on previously published stories:
\begin{quote}
    In the case that I start following long-running content after many episodes have already been published, I think it would be exciting if I were able to refer to other people's speculations or discussions. However, considering the possibility of being spoiled by searching them on the Internet, it is not possible. (P4)
\end{quote}
\begin{quote}
    I would like to take a look at speculations for each episode in order to savor it. But given the risk of spoilers, it's hard to do. (P3)
\end{quote}
These comments suggest that it would be important to avoid spoilers in developing support for catching-up users by leveraging online speculative discussions.

\section{Tweets Analysis for Characterizing Online Reaction- and Speculation-Based Media Consumption}
\label{sec:tweets-analysis}

In \secref{introduction} and \secref{interaction-techniques}, we discussed the possibility of providing computational support for catching-up users by presenting tweets that share immediate reactions.
At the same time, our semi-structured interviews confirmed the positive aspects of engaging with the online speculative discussions, which implies another emerging approach for supporting catching-up users.
In particular, given that the effect of deepening understandings of media content would be specific to speculation (\secref{interview-understanding}), it is worth to explore both approaches.

To develop computational support for catching-up users via these approaches, we need to construct a deeper understanding based on actual data regarding both of immediate reactions and speculations.
For this purpose, we conducted tweets analysis in regard to reaction- and speculation-based media consumption since Schirra et al. \cite{DBLP:conf/chi/SchirraSB14} and the participants in \secref{interview-findings} suggested the use of Twitter as a platform to relate to those activities.
This analysis is crucial to develop computational support because the approaches would not be feasible without the tweet data to be presented.
In addition, the characteristics of such tweets can provide implications for designing computational support, which we discuss later in \secref{user-experiment}.
We thus reviewed two examples of media content and analyzed users' behaviors quantitatively and qualitatively using publicly available data.

\subsection{Data Collection from Twitter}
\label{sec:tweets-data-collection}

As Twitter has over 330 million monthly active users \cite{Washington_2019}, previous research has used Twitter data to understand social behaviors in diverse areas, including politics \cite{bruns2011ausvotes}, crises \cite{bruns2012qldfloods}, and TV programs \cite{deller2011twittering}.
For example, as mentioned in \secref{internet-based-social-interactions}, some studies examined live-tweeting during TV programs by classifying types of content posted \cite{FM3368} or analyzing user motivations \cite{DBLP:conf/chi/SchirraSB14}.
More specifically, Schirra et al. \cite{DBLP:conf/chi/SchirraSB14} presented a time plot of tweeting activities during scheduled airtimes and suggested the existence of ``have-to-tweet-about'' moments, represented by triggers in the story that tempt users to tweet.
Our approach is similar in that we collected tweets and analyzed them both quantitatively and qualitatively.
However, we targeted not only on reaction-related activities that these studies dealt with but also on speculation-related activities.

For the purpose of comparing the two activities, we used hashtag searches to gather tweets around the airtimes of the content.
We first determined hashtags corresponding to two categories: (1) tweets involving speculation about a target media content (\textit{speculation-tweets}), and (2) other tweets about the same content (\textit{non-speculation-tweets}).
Here, we reviewed tweets containing the title of the target media content and manually extracted popular hashtags for each category, whose process is detailed in the following sections.
We note that this is based on our anticipation that \textit{speculation-tweets} would contain specific words or hashtags, such as ``考察'', as suggested in \secref{interview-understanding}.
In contrast, most \textit{non-speculation-tweets} would reflect the reaction-related activities while there would be a few tweets used for other purposes such as pre-broadcast advertising by the official marketing account or the actors of the TV show.

For each hashtag found, we collected corresponding tweets that had been posted during a specified period around the broadcast date of certain episodes.
We also took into account that tweets could contain multiple hashtags, leading search results for different hashtags to include the same tweets.
To ensure an accurate count, we thus removed duplicated tweets using the tweet post ID after aggregating the results of each search.
In addition, if a tweet contained hashtags for both \textit{speculation-tweets} and \textit{non-speculation-tweets}, then we assumed the tweet was intended to share speculation from our observation and classified it as a \textit{speculation-tweet}.
We used the collected data to analyze the characteristics of tweeting behavior related to speculation, including the timing, length, and content of tweets.

\subsection{Case Study: \textit{Your Turn to Kill} (あなたの番です)}
\label{sec:tweets-anaban}

We first examined tweets about the Japanese TV series, \textit{Your Turn to Kill} (あなたの番です), a fictional narrative set in an apartment building where serial murders occur.
As described in \secref{introduction}, this series was designed to encourage speculation on social networking services, where viewers often posted their opinions and discussed their predictions about the next victim or the true culprit.

To determine hashtags for data collection, we searched tweets using the title of the TV show as a hashtag.
Our initial results indicated that, in addition to the title, its simple abbreviation, ``\#あな番,'' was frequently used in the tweets.
Since we observed these two hashtags were widely used for sharing immediate reactions, we selected them for \textit{non-speculation-tweets}.
We next found that \textit{speculation-tweets} incorporated hashtags that appended the word ``考察'' [meaning ``speculation'' in Japanese] to the two hashtags (``\#あなたの番です考察'' or ``\#あな番考察'').
Interestingly, we also discovered that a specific hashtag had been coined for posting speculation about this TV series: ``\#オラウータンタイム'' [meaning ``time of orangutan'' in Japanese].
The use of this hashtag is attributable to the main character's frequent use of this term when speculating about solutions to the mysteries he confronts.\footnote{According to online speculative discussions, this term seemed to come from ``The Murders in the Rue Morgue'' by Edgar Allan Poe.}
It is notable that the viewers also reused this term to present their own speculations.

The TV series was broadcast weekly between April 14, 2019, and September 8, 2019.
We selected one episode from the middle of the series and a second episode from the later in the series, which were broadcast on June 9 from 22:30--23:30 JST and September 1 from 22:30--23:30 JST, respectively. 
To observe viewers' behavior around these airtimes, we collected tweets approximately one day before and after each airtime, i.e., from June 9 at 0:00 JST to June 10 at 23:59 JST and from September 1 from 0:00 JST to September 2 at 23:59 JST. 
\tabref{anaban-dataset} summarizes the selected hashtags and the total number of tweets for both \textit{speculation-tweets} and \textit{non-speculation-tweets} during each period.

\begin{table}[tb]
    \caption{Hashtags selected and the number of tweets collected for \textit{speculation-tweets} and \textit{non-speculation-tweets} about \textit{Your Turn to Kill} (あなたの番です).}
    \label{tbl:anaban-dataset}
    \scalebox{0.8}{
    \renewcommand{\arraystretch}{1.05}
    \begin{tabular}{l@{\hspace{1.5em}}c@{\hspace{1.5em}}rr} \toprule
    \multicolumn{1}{c}{\multirow{2.3}{*}{Tweet category}} & \multirow{2.3}{*}{Hashtags}                      & \multicolumn{2}{c}{Number of total tweets during}                                  \\ \cmidrule{3-4}
                                                          &                                                  & Jun. 9 -- Jun. 10                       & Sep. 1 -- Sep. 2                         \\ \midrule
    \multirow{3}{*}{\textit{speculation-tweets}}          & \#あなたの番です考察 \ [title + ``speculation''] & \multirow{3}{*}{\nAnabanSpeculationMid} & \multirow{3}{*}{\nAnabanSpeculationLate} \\
                                                          & \#あな番考察 \ [abbr. title + ``speculation'']   &                                         &                                          \\
                                                          & \#オラウータンタイム \ [``time of orangutan'']   &                                         &                                          \\ \midrule[0.2pt]
    \multirow{2}{*}{\textit{non-speculation-tweets}}      & \#あなたの番です \ [title]                       & \multirow{2}{*}{\nAnabanContentMid}     & \multirow{2}{*}{\nAnabanContentLate}     \\
                                                          & \#あな番 \ [abbr. title]                         &                                         &                                          \\ \bottomrule
    \end{tabular}
    }
\end{table}

In the middle episode period, there were \nAnabanSpeculationMid~and \nAnabanContentMid~total \textit{speculation-tweets} and \textit{non-speculation-tweets}, respectively.
During the late episode period, these numbers increased to \nAnabanSpeculationLate~and \nAnabanContentLate, respectively, as the program became more popular with a greater number of fans involved in online communication.
The ratio of \textit{speculation-tweets} to \textit{non-speculation-tweets} also significantly increased, from 6.7\% in the middle episode period to 10.2\% in the late episode period---that is, more \textit{speculation-tweets} posted toward the end of the TV series as the climax of the story approached.

\begin{figure}[tb]
    \begin{minipage}{0.8\hsize}
        \begin{center}
            \includegraphics[width=0.99\textwidth]{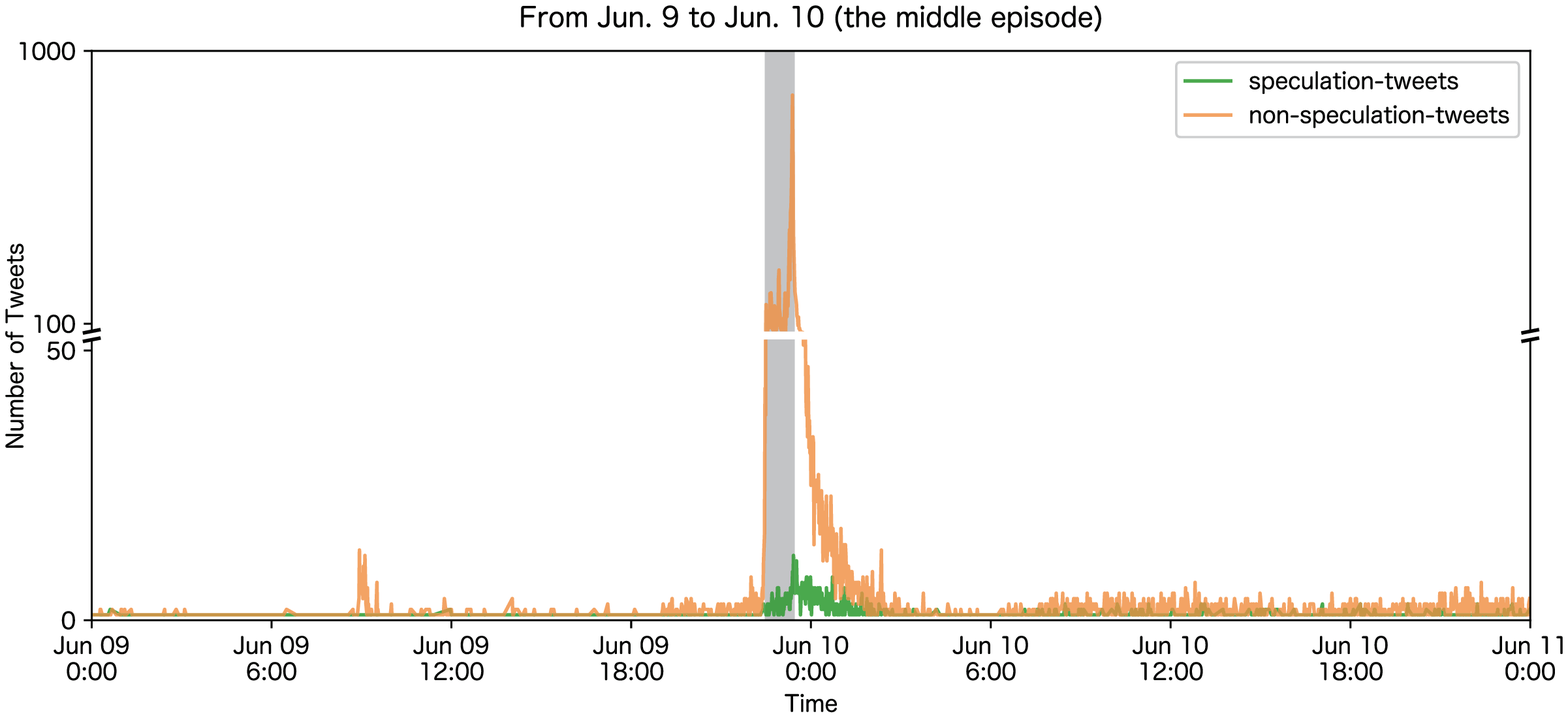}
        \end{center}
    \end{minipage}
    \begin{minipage}{0.8\hsize}
        \begin{center}
            \includegraphics[width=0.99\textwidth]{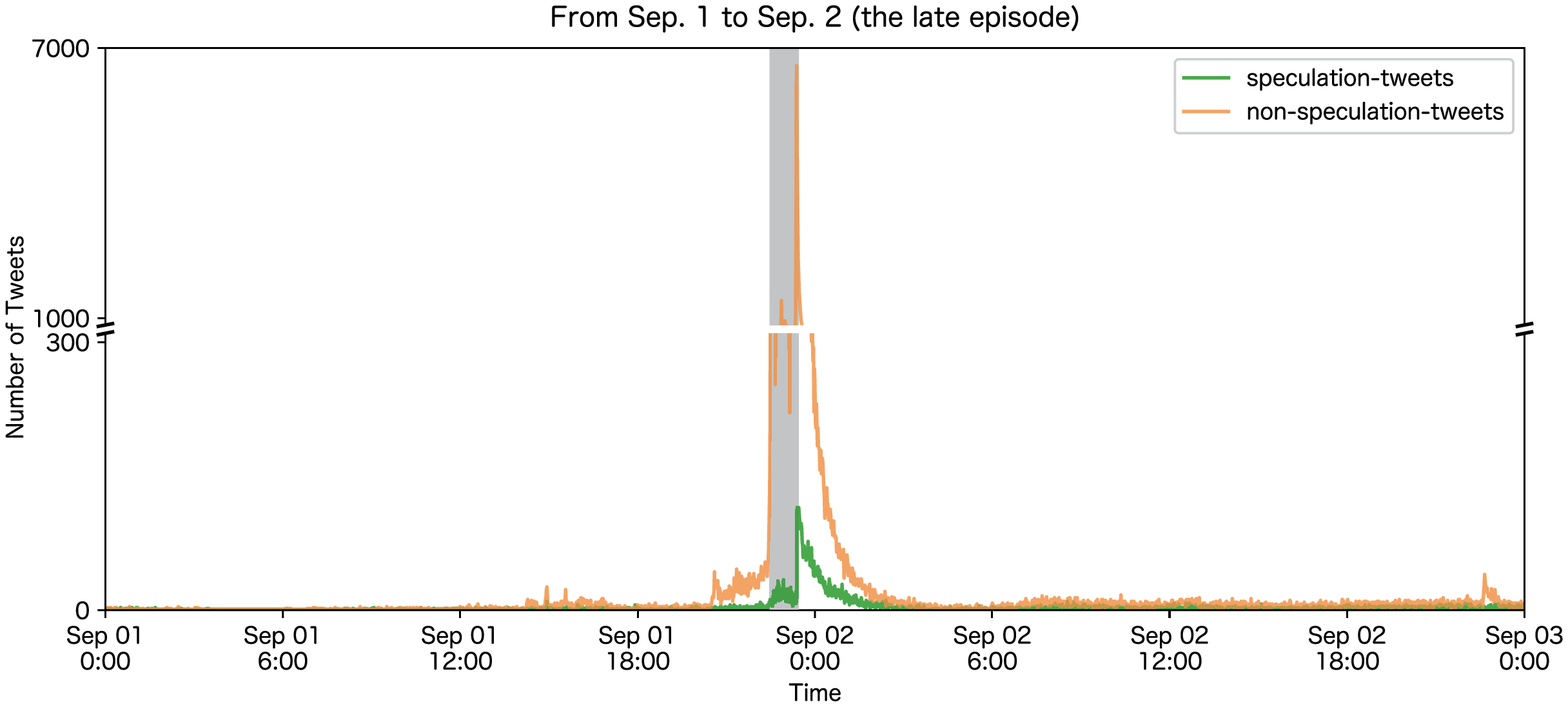}
        \end{center}
    \end{minipage}
    \caption{Time plots of the number of \textit{speculation-tweets} and \textit{non-speculation-tweets} within the two observed periods (top: the middle episode, bottom: the late episode) for \textit{Your Turn to Kill} (あなたの番です). Gray backgrounds denote each episode's airtime.}
    \label{fig:anaban_number_of_tweets}
\end{figure}

A time plot of the volume of tweets from this dataset is presented in \figref{anaban_number_of_tweets}.
Tweets were counted on a minute basis, with units smaller than minutes are rounded down.
Similar to the result of the previous research on live-tweeters \cite{DBLP:conf/chi/SchirraSB14}, peak volume was aligned with the episode's airtime for both \textit{speculation-tweets} and \textit{non-speculation-tweets}.

\begin{figure}[tb]
    \begin{minipage}{0.8\hsize}
        \begin{center}
            \includegraphics[width=0.99\textwidth]{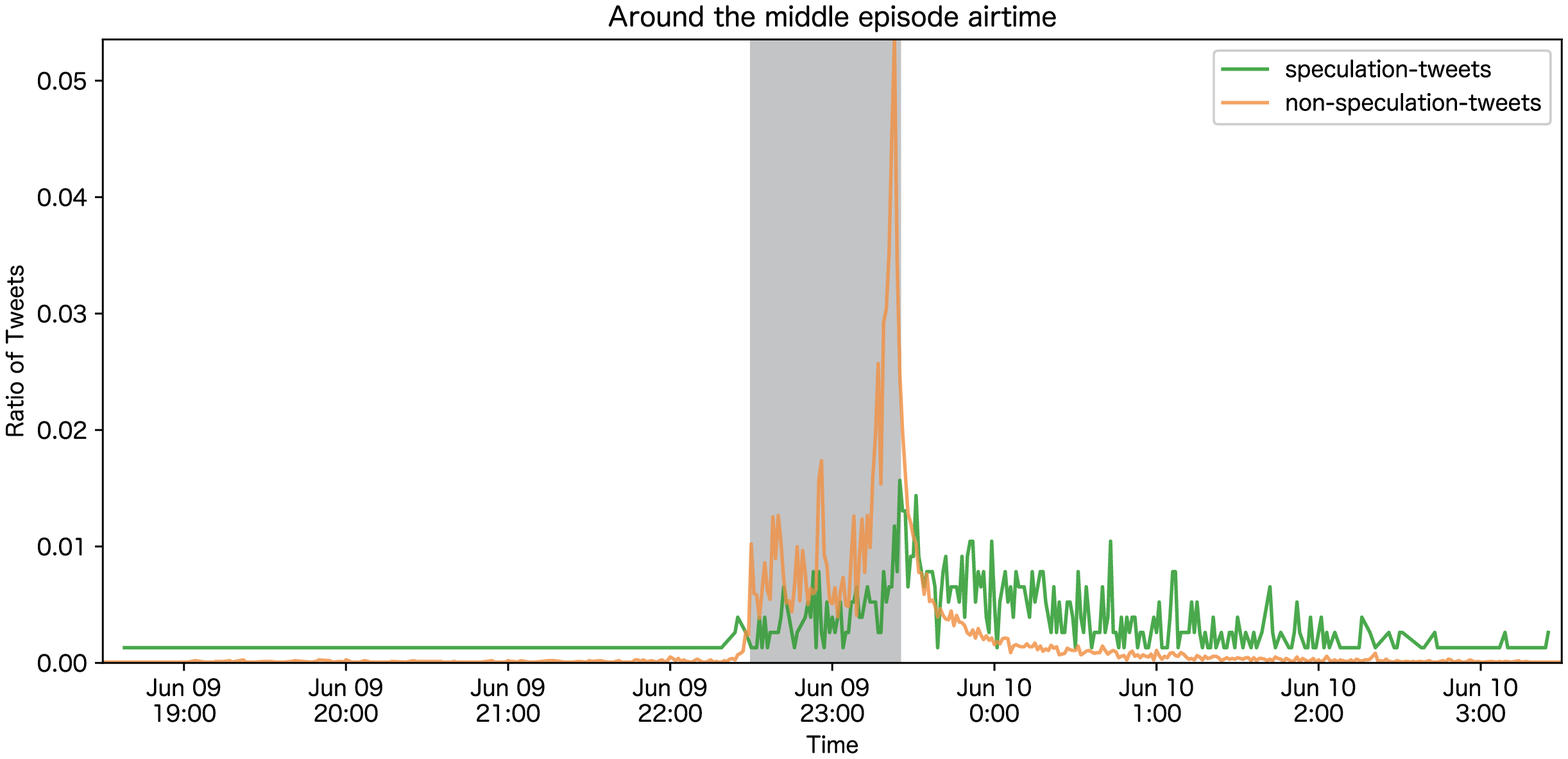}
        \end{center}
    \end{minipage}
    \begin{minipage}{0.8\hsize}
        \begin{center}
            \includegraphics[width=0.99\textwidth]{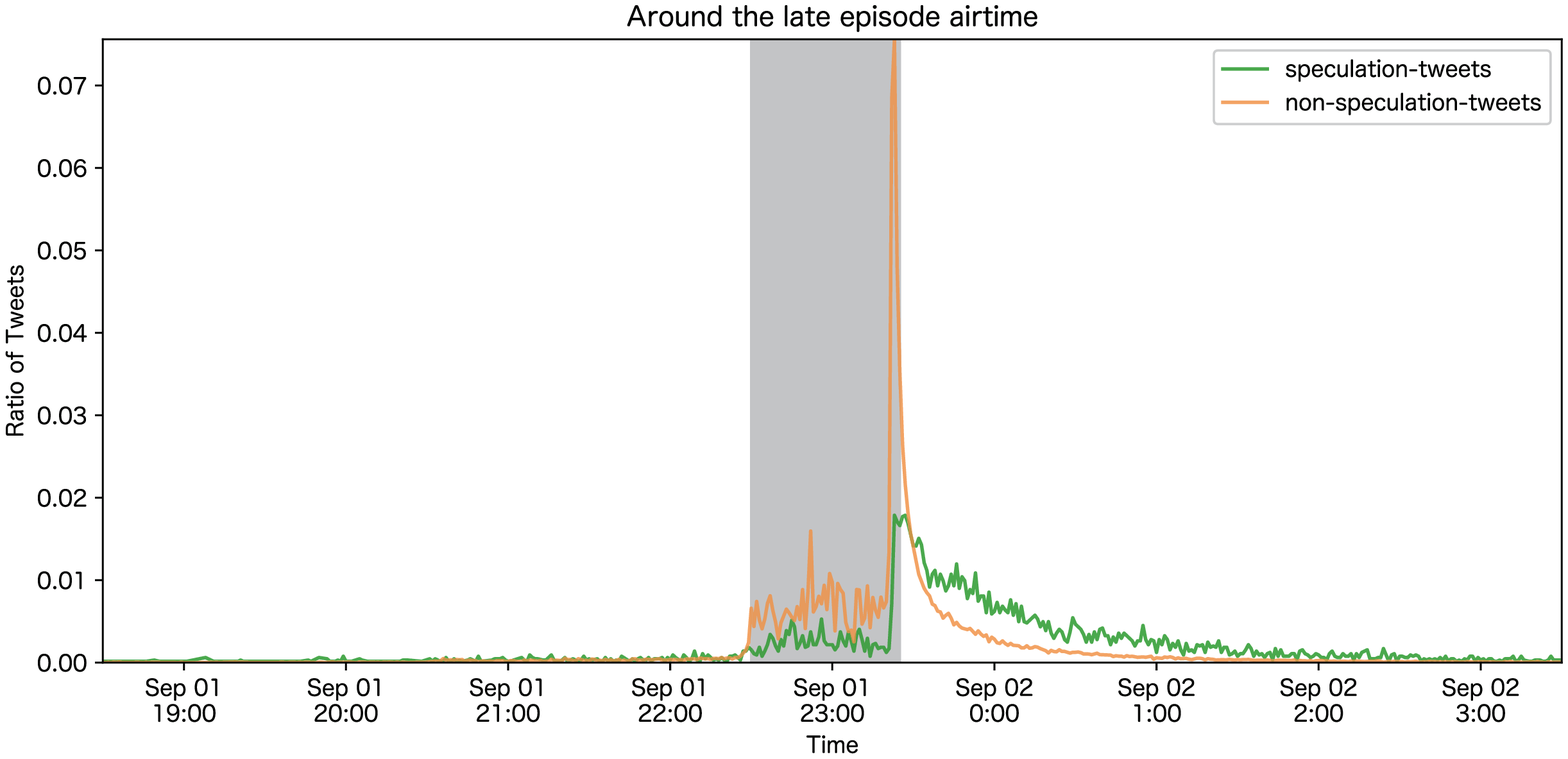}
        \end{center}
    \end{minipage}
    \caption{Time plots of the relative volume of \textit{speculation-tweets} and \textit{non-speculation-tweets} during the two observed periods (top: the middle episode, bottom: the late episode) for \textit{Your Turn to Kill} (あなたの番です). Gray backgrounds denote the airtime of each episode.}
    \label{fig:anaban_ratio_of_tweets}
\end{figure}

To further analyze the timing of users' involvement in the speculation, we created a time plot of the normalized volume of tweets around the episode's airtime (18:30--27:30), calculated by dividing by the total number of tweets in each category during the period.
The difference in behavior between \textit{speculation-tweets} and \textit{non-speculation-tweets} is illustrated in \figref{anaban_ratio_of_tweets}.
While the peaks of both \textit{speculation-tweets} and \textit{non-speculation-tweets} were concentrated near the end of each airtime, we observed lingering segments of \textit{speculation-tweets} within a few hours after the end of the broadcast.
This suggests that \textit{speculation-tweets} are likely to appear around and after the end of the program, which can be attributed to viewers' motivation to post forecasts about what will happen in the next episode, rather than sharing immediate reactions.
We infer this trend would be a unique property of media consumption related to speculation, in comparison with previously observed consumption experiences centered on immediate reactions \cite{DBLP:conf/chi/SchirraSB14}.

Next, we investigated the contents of the tweets in the dataset to characterize qualitative perspectives.
In addition to the observation that text information in \textit{speculation-tweets} was related to predicting future story developments compared to that in \textit{non-speculation-tweets}, we found that \textit{speculation-tweets} often contained hyperlinks to external content---such as images, other tweets, blogs and other articles, and YouTube videos.
To confirm this, we classified the tweets based on pattern-matching of the contained URL and calculated the ratio of tweets in each hyperlink type to the total number of tweets.
As shown in \tabref{anaban_link}, the ratio of containing hyperlinks in \textit{speculation-tweets} is nearly double that of \textit{non-speculation-tweets} for each type.

\begin{table}[tb]
    \caption{Ratio of tweets containing hyperlinks by the type of its destination for \textit{Your Turn to Kill} (あなたの番です).}
    \label{tbl:anaban_link}
    \scalebox{0.8}{
    \begin{tabular}{ll@{\hspace{1em}}cccc@{\hspace{1em}}c} \toprule
    \multicolumn{1}{c}{\multirow{2.3}{*}{Period}} & \multicolumn{1}{c}{\multirow{2.3}{*}{Tweet category}} & \multicolumn{4}{c}{Hyperlink types}                               & \multirow{2.3}{*}{Total} \\ \cmidrule(l{0em}r{1em}){3-6}
                                                  &                                                       & Images  & Other tweets & YouTube videos & Blogs \& other articles &                          \\ \midrule
    \multirow{2}{*}{The middle episode}           & \textit{speculation-tweets}                           & 13.89\% & 3.11\%       & 0.64\%         & 5.67\%                  & 23.30\%                  \\
                                                  & \textit{non-speculation-tweets}                       & 5.50\%  & 1.52\%       & 0.08\%         & 2.86\%                  & 9.96\%                   \\ \midrule[0.2pt]
    \multirow{2}{*}{The late episode}             & \textit{speculation-tweets}                           & 18.68\% & 3.16\%       & 0.50\%         & 8.98\%                  & 31.32\%                  \\
                                                  & \textit{non-speculation-tweets}                       & 9.97\%  & 1.08\%       & 0.13\%         & 1.46\%                  & 12.64\%                  \\ \midrule[0.2pt]
    \multirow{2}{*}{Total}                        & \textit{speculation-tweets}                           & 18.14\% & 3.16\%       & 0.52\%         & 8.61\%                  & 30.42\%                  \\
                                                  & \textit{non-speculation-tweets}                       & 9.27\%  & 1.15\%       & 0.12\%         & 1.68\%                  & 12.22\%                  \\ \bottomrule
    \end{tabular}
    }
\end{table}

\begin{figure}[tb]
    \begin{center}
        \includegraphics[width=0.3\textwidth]{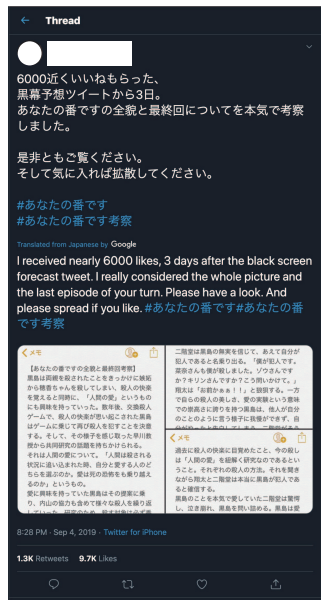}
    \end{center}
    \caption{A tweet which includes a screenshot of a note-taking app for the purpose of sharing long sentences of speculation.\protect\footnotemark}
    \label{fig:screenshot-tweet}
\end{figure}

The images in \textit{non-speculation-tweets} were observed to be often pictures of the actors or actresses on the show, accompanied by comments such as ``Her acting is amazing'' or ``He is cool!''
In contrast, when sorting tweets by number of likes, we observed a wide variety of \textit{speculation-tweets} incorporated different types of images.
In particular, we found a number of screenshots of note-taking apps containing extensive speculations, which correspond to the comments by P7 in \secref{interview-understanding}.
An example of one such tweet is shown in \figref{screenshot-tweet}.
This suggests a potential emerging usage of Twitter as a tool to deliver lengthy speculations.

\footnotetext{This is taken from \url{https://twitter.com/WXkKXnSZVF0zObE/status/1169210679619813376} (Accessed: January 1, 2020).}

Similar behaviors were observed for blogs and other articles and YouTube videos.
The linked content in \textit{speculation-tweets} often presented opinions and predictions about the next victim or the true culprit, while links in \textit{non-speculation-tweets} typically included content with which fans were pleased with, such as costume design of the show's actors and actresses.
Moreover, \textit{speculation-tweets} were often quoted in other tweets with comments such as ``I'm just amazed by this reasoning. Certainly, this idea can explain that scene,'' ``I don't agree with this because this contradicts his alibi,'' or ``I totally didn't notice this point. I need to watch the episode again.''
Another unique usage of Twitter was the creation of collection threads of \textit{speculation-tweets} using Twitter's threading function, with additional meta-speculation based on a comparison of the listed tweets.

\begin{figure}[tb]
    \begin{center}
        \begin{subfigure}{0.4\textwidth}
            \includegraphics[width=0.99\textwidth]{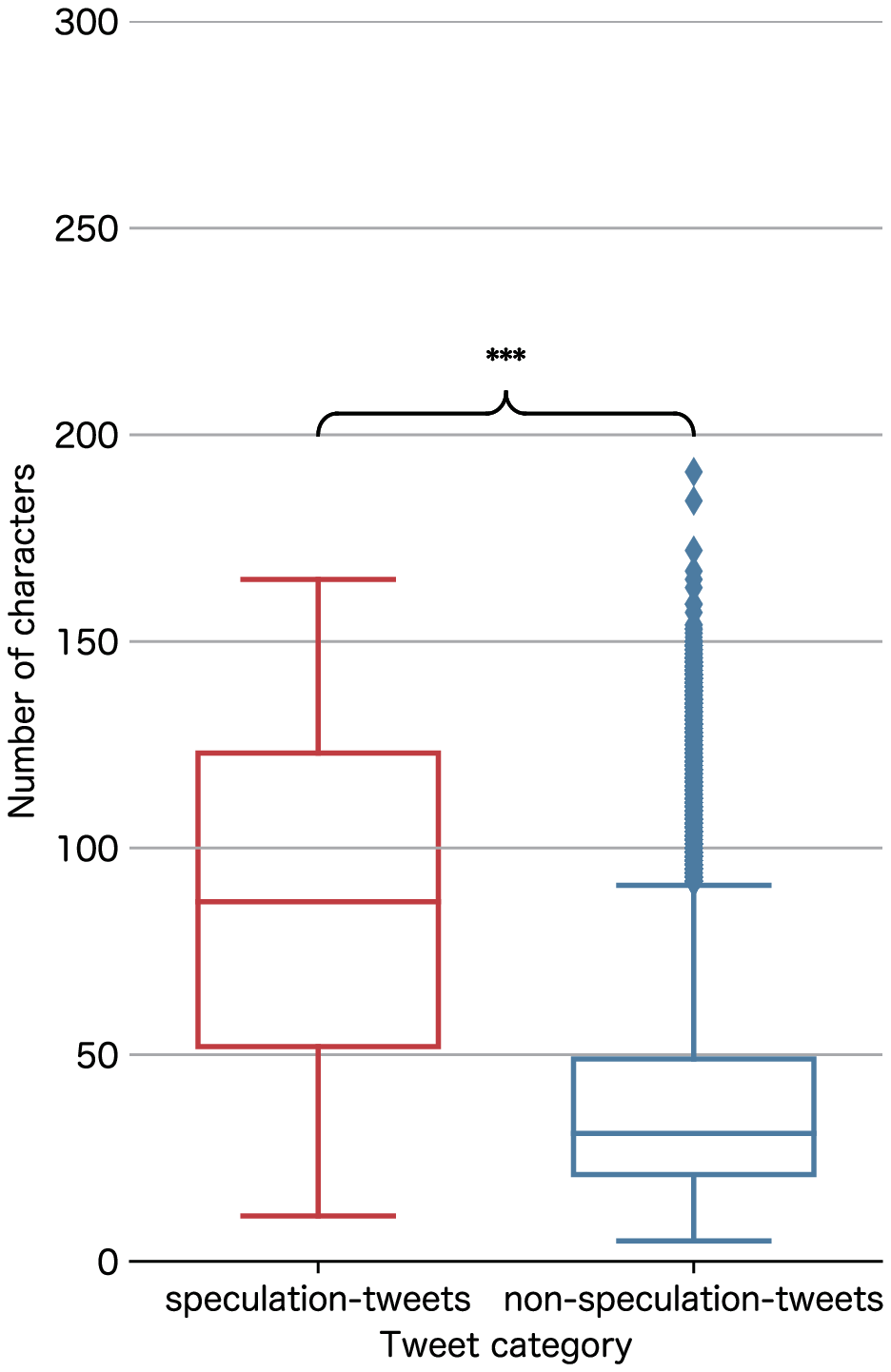}
        \end{subfigure}
        \begin{subfigure}{0.4\textwidth}
            \includegraphics[width=0.99\textwidth]{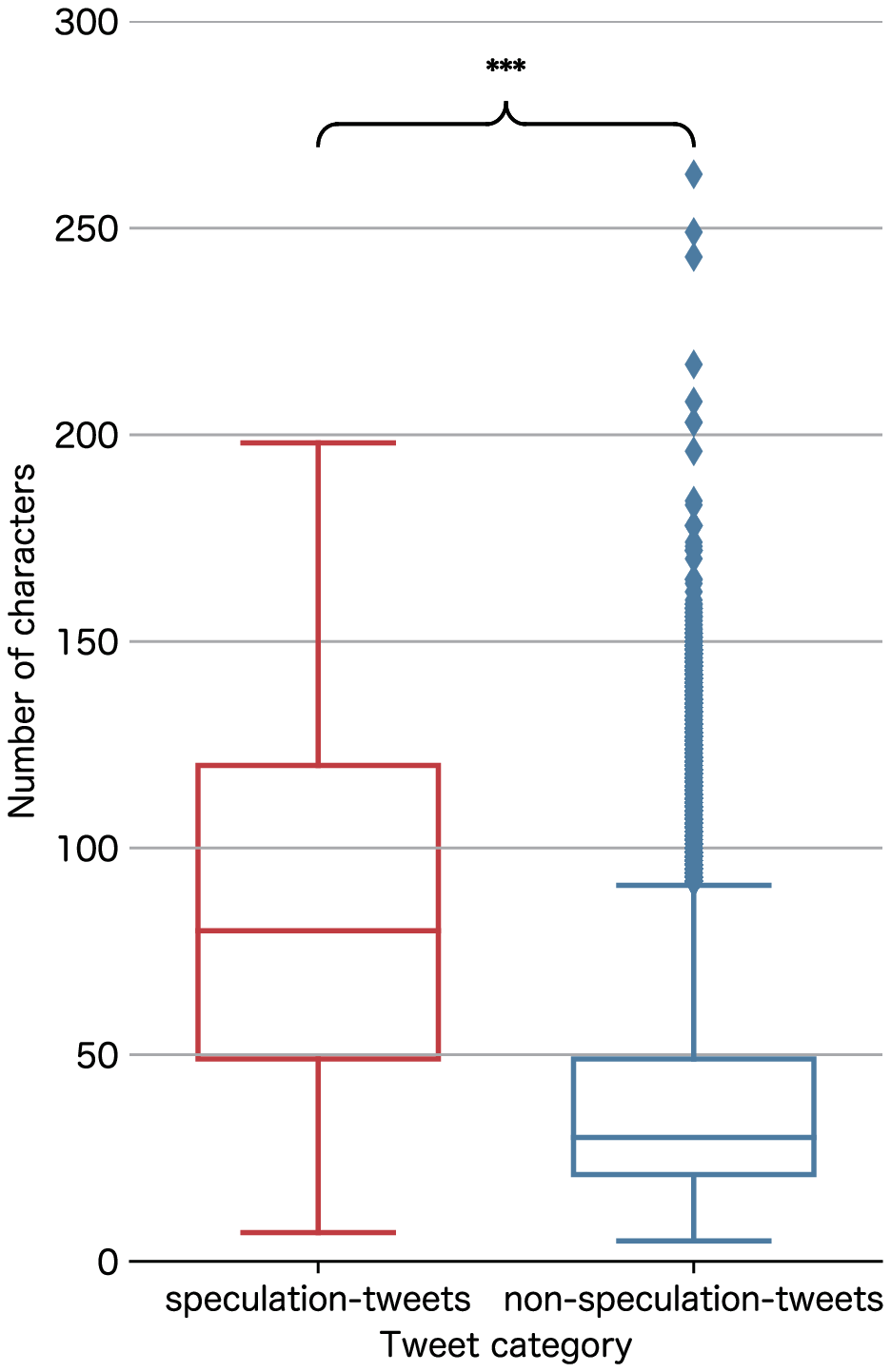}
        \end{subfigure}
    \end{center}
    \caption{Number of characters in \textit{speculation-tweets} and \textit{non-speculation-tweets} during the two observed periods (top: the middle episode, bottom: the late episode) for \textit{Your Turn to Kill} (あなたの番です). \textit{Speculation-tweets} contained significantly more characters than \textit{non-speculation-tweets} ($ p < 0.001 $).}
    \label{fig:anaban_number_of_characters}
\end{figure}

Finally, to explore the quantitative difference in the information that users shared in \textit{speculation-tweets} versus \textit{non-speculation-tweets}, we compared the number of characters contained in each type of tweets.
Here, we excluded the tweets with hyperlinks in order to count the number of characters in text without URLs.
As a result, the number of characters in \textit{speculation-tweets} was significantly larger than that in \textit{non-speculation-tweets} (\figref{anaban_number_of_characters}), which suggests a distinct user behavior of sharing lengthier, speculative thoughts and opinions that differ from immediate responses to the program.

These results suggest the unique characteristics of \textit{speculation-tweets} in comparison with those of \textit{non-speculation-tweets}, such as their lingering occurrence after the airtime, longer text length, and tendency to contain hyperlinks, which were not covered in previous studies.
Therefore, we would be required to adopt different strategies for providing opportunities for catching-up users to relate to immediate reactions and speculations.
To ensure that this distinction is not specific to \textit{Your Turn to Kill} (あなたの番です), we further investigated speculation-based media consumption with different media content.

\subsection{Case Study: \textit{Ship of Theseus} (テセウスの船)}
\label{sec:tweets-theseus}

Next, we analyzed another recent TV series, \textit{Ship of Theseus} (テセウスの船), in the same manner.
This Japanese TV series is set in a science-fiction world where the lead character travels into the past to find clues about unresolved mysteries related to his father's dishonor.
The series was broadcast weekly from January 19, 2020, to March 22, 2020.
As a tightly plotted story with many episodes foreshadowing upcoming events, we anticipated that a large number of viewers would participate in speculation about upcoming content on the Internet.

Similar to the case of \textit{Your Turn to Kill} (あなたの番です), we first searched tweets containing the title of the TV show to determine the hashtags for data collection.
As a result, we found a simple abbreviation of the title, ``\#テセウス'' [Theseus], was also frequently used in the tweets.
Since we observed these hashtags were widely used for sharing immediate reactions, we selected them for \textit{non-speculation-tweets}.
Also, we found that \textit{speculation-tweets} incorporated hashtags that appended the word ``考察'' [meaning ``speculation'' in Japanese] to the two hashtags used for \textit{non-speculation-tweets}. 
Note that these observations were consistent with the previous case (\secref{tweets-anaban}).

\tabref{theseus-dataset} presents the hashtags used for data collection, the data collection period, and the number of \textit{speculation-tweets} and \textit{non-speculation-tweets}.
We selected one late episode, which was broadcast on March 15 from 21:00 to 22:00 JST, and collected tweets from approximately one day before and after the episode aired.

\begin{table}[tb]
    \caption{Hashtags selected and the number of tweets collected for \textit{speculation-tweets} and \textit{non-speculation-tweets} about \textit{Ship of Theseus} (テセウスの船).}
    \label{tbl:theseus-dataset}
    \scalebox{0.8}{
    \renewcommand{\arraystretch}{1.05}
    \begin{tabular}{l@{\hspace{1em}}c@{\hspace{1em}}r} \toprule
    \multicolumn{1}{c}{\multirow{2}{*}{Tweet category}} & \multirow{2}{*}{Hashtags}                        & \multicolumn{1}{c}{Number of total tweets during} \\
                                                        &                                                  & \multicolumn{1}{c}{Mar. 15 -- Mar. 16}            \\ \midrule
    \multirow{2}{*}{\textit{speculation-tweets}}        & \#テセウスの船考察 \ [title + ``speculation'']   & \multirow{2}{*}{\nTheseusSpeculation}             \\
                                                        & \#テセウス考察 \ [abbr. title + ``speculation''] &                                                   \\ \midrule[0.2pt]
    \multirow{2}{*}{\textit{non-speculation-tweets}}    & \#テセウスの船 \ [title]                         & \multirow{2}{*}{\nTheseusContent}                 \\
                                                        & \#テセウス \ [abbr. title]                       &                                                   \\ \bottomrule
    \end{tabular}
    }
\end{table}

\begin{figure}[tb]
    \begin{minipage}{0.8\hsize}
        \begin{center}
            \includegraphics[width=0.99\textwidth]{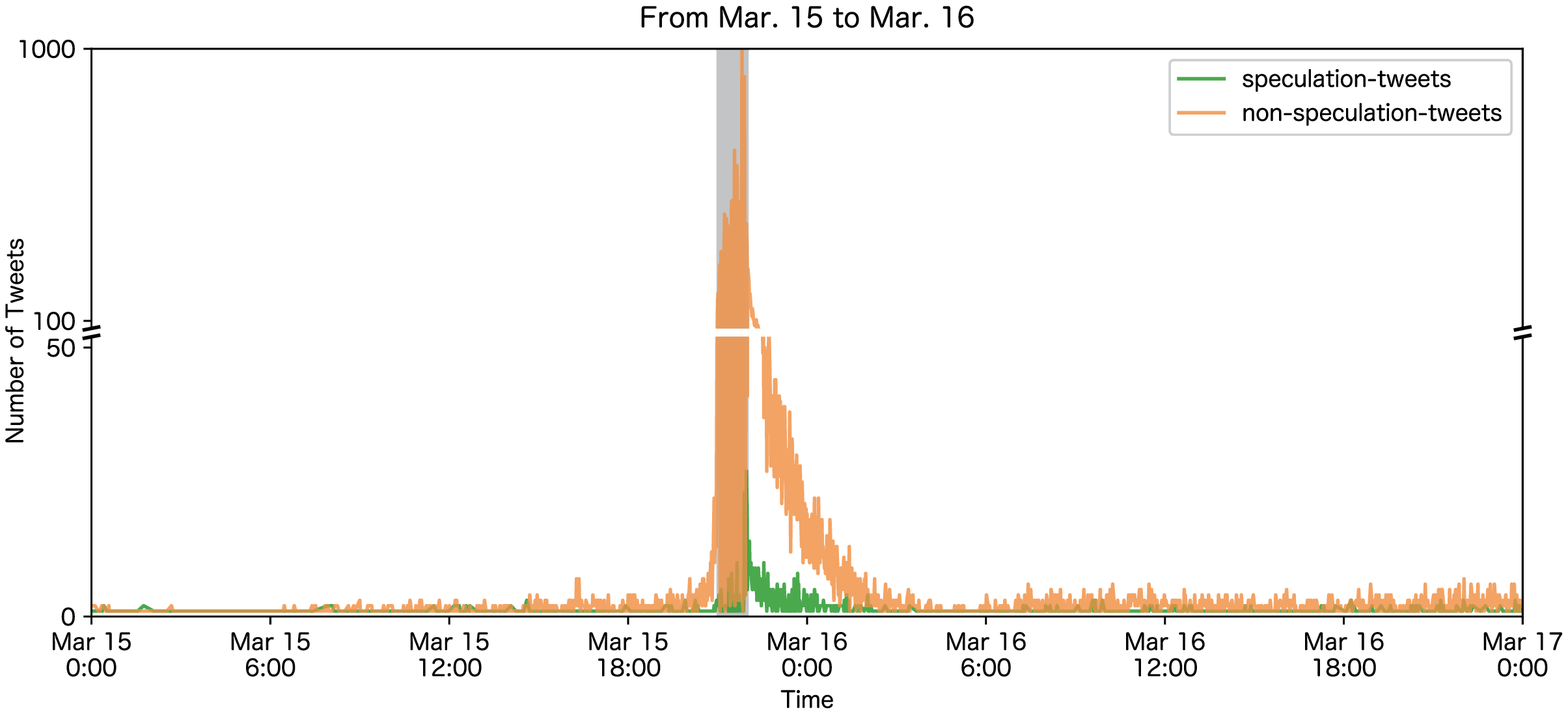}
        \end{center}
    \end{minipage}
    \begin{minipage}{0.8\hsize}
        \begin{center}
            \includegraphics[width=0.99\textwidth]{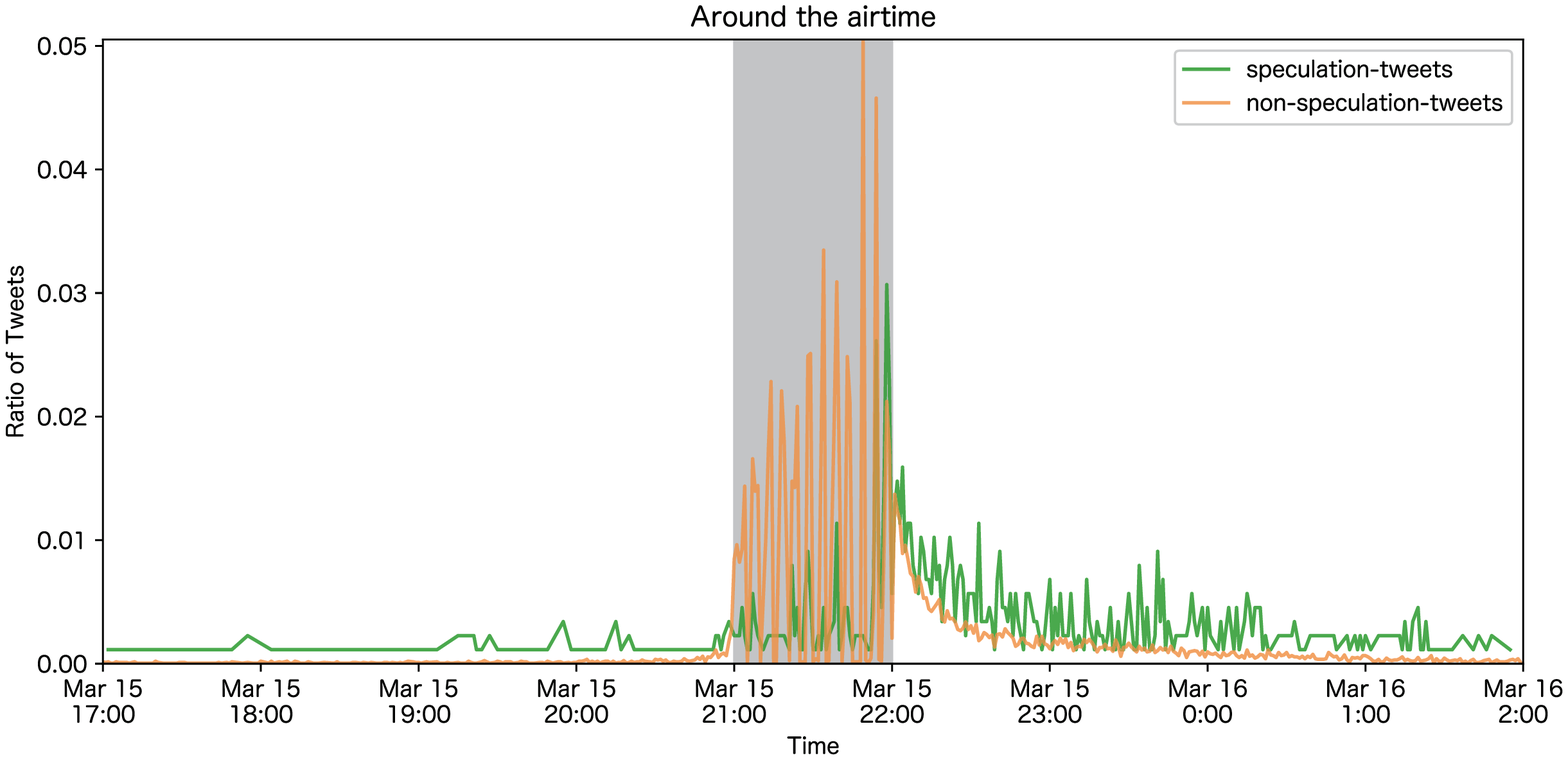}
        \end{center}
    \end{minipage}
    \caption{Time plots of the volumes of \textit{speculation-tweets} and \textit{non-speculation-tweets} (top: the quantity, bottom: the relative volume) for \textit{Ship of Theseus} (テセウスの船). Gray backgrounds denote the airtime.}
    \label{fig:theseus_tweets_timeplot}
\end{figure}

\figref{theseus_tweets_timeplot} presents a time plot of the volume and ratio of the collected tweets on a minute basis.
Here, we observed a lingering segment of \textit{speculation-tweets} after the airtime, similar to those identified in \figref{anaban_number_of_tweets} and \figref{anaban_ratio_of_tweets}.

A similar commonality was also presented in the ratio of tweets containing hyperlinks (\tabref{theseus_link}).
The ratio for \textit{Ship of Theseus} (テセウスの船) was similar to that for \textit{Your Turn to Kill} (あなたの番です).
In particular, tweets containing hyperlinks to blogs and other articles were much more frequent in \textit{speculation-tweets}.

\begin{table}[tb]
    \caption{Ratio of tweets containing hyperlinks by the types of destination for \textit{Ship of Theseus} (テセウスの船).}
    \label{tbl:theseus_link}
    \scalebox{0.8}{
    \begin{tabular}{l@{\hspace{1em}}cccc@{\hspace{1em}}c} \toprule
    \multicolumn{1}{c}{\multirow{2.3}{*}{Tweet category}} & \multicolumn{4}{c}{Hyperlink types}                               & \multirow{2.3}{*}{Total} \\ \cmidrule(l{0em}r{1em}){2-5}
                                                          & Images  & Other tweets & YouTube videos & Blogs \& other articles &                          \\ \midrule
    \textit{speculation-tweets}                           & 15.65\% & 4.18\%       & 1.32\%         & 21.38\%                 & 42.53\%                  \\
    \textit{non-speculation-tweets}                               & 8.75\%  & 1.69\%       & 0.28\%         & 3.13\%                  & 13.85\%                  \\ \bottomrule
    \end{tabular}
    }
\end{table}

\begin{figure}[tb]
    \begin{center}
        \includegraphics[width=0.4\textwidth]{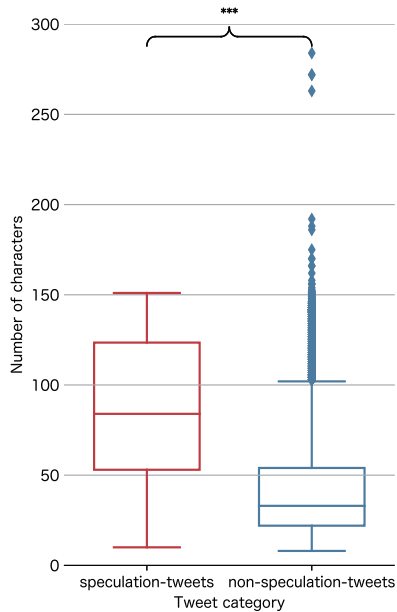}
    \end{center}
    \caption{Number of characters in \textit{speculation-tweets} and \textit{non-speculation-tweets} about \textit{Ship of Theseus} (テセウスの船). \textit{Speculation-tweets} contained significantly more characters than \textit{non-speculation-tweets} ($ p < 0.001 $).}
    \label{fig:theseus_number_of_characters}
\end{figure}

Furthermore, we compared the number of characters in a tweet in the same manner as \figref{anaban_number_of_characters}, which yielded a result that met our expectations---namely, significantly more characters were observed in \textit{speculation-tweets} than in \textit{non-speculation-tweets} (\figref{theseus_number_of_characters}).

These results support the difference in the characteristics of \textit{speculation-tweets} versus \textit{non-speculation-tweets} described in \secref{tweets-anaban} and, in particular, illustrate the unique aspects of sharing speculations about serial content.
In other words, in the context of developing computational support for catching-up users, different strategies should be adopted for providing opportunities to relate to immediate reactions or speculations.

\section{User Experiment to Evaluate Support for Catching-Up Users}
\label{sec:user-experiment}

Until this point, our semi-structured interviews illustrated that engaging with speculative discussions not only gives a sense of being connected with others but also deepens understandings of media content.
Furthermore, our tweets analysis revealed distinctive characteristics in the tweet data of reaction- and speculation-based media consumption, which suggested the need for adopting different approaches to present each data to catching-up users.

To lay the groundwork for discussing the possibility of computational support for catching-up users, we prepared prototypes based on the two different approaches.
We compared their effectiveness by replicating the situation of catching-up users in a user experiment.
The first prototype provides immediate reactions while watching, and the second offers speculative discussions after each episode.
In this section, we describe our procedure and the results of the user experiment.

\subsection{Materials}

To replicate the situation of catching-up users, we needed to select a TV series to present to participants.
We chose \textit{Your Turn to Kill} (あなたの番です) because its suitability for speculation was confirmed by the number of \textit{speculation-tweets} (see \secref{tweets-anaban}).
In addition, since we had collected tweets related to this TV show in \secref{tweets-data-collection}, we were able to use our previously acquired data to construct the prototypes. 

\subsection{Participants}

Our experiment involved 18 participants aged between 20 and 52, of whom five were female.
They were recruited via word of mouth and online communication in the same manner as described in \secref{interview-participants}.
In order to be eligible, participants had to speak Japanese, since a Japanese-language TV series was used in this experiment.
At the time of recruitment, all participants self-reported that they had never watched \textit{Your Turn to Kill} (あなたの番です) and were not familiar with the story.

\subsection{Prototypes}
\label{sec:experiment-prototypes}

To explore possibilities of computational support for catching-up users, we prepared two different prototypes: live-tweets and speculation display.
The first displayed tweets posted during the original broadcast in relation to playback time within the episode, similar to Danmaku \cite{Wu:2019:DNP:3340675.3329485}, as we mentioned in \secref{introduction} and \secref{interaction-techniques}.
Since we confirmed that tweets associating with such immediate reactions were concentrated around the airtime (\secref{tweets-analysis}), this design would be reasonable.
In addition, given the effect of Danmaku \cite{DBLP:journals/chb/LimHKB15} and Twitter-based social viewing experience \cite{DBLP:conf/chi/SchirraSB14}, we could expect that participants would appreciate the immediate reactions when using this prototype while watching the episode.

The second prototype was intended to induce the positive effects of engaging with speculative discussions described in \secref{interview-findings}.
Here, since speculative discussions often have long sentences or hyperlinks to external content (\secref{tweets-analysis}), a Danmaku-based interface would not be optimal.
We thus constructed a unique web page for each episode that displayed \textit{speculation-tweets} posted between the broadcast of that episode and the following episode.
This post-time-based data collection is intended to eliminate the risk of spoilers, as discussed in \secref{interview-spoilers}.
The tweets were randomly arranged on the web page.
We considered that participants interact with this page after they finished watching each episode, expecting that they would engage with speculation.
This design is analogous to our findings in \secref{tweets-analysis}; speculative discussions on Twitter become active at the end of and after the airtime.

\begin{figure}[tb]
    \begin{center}
        \includegraphics[width=0.65\textwidth]{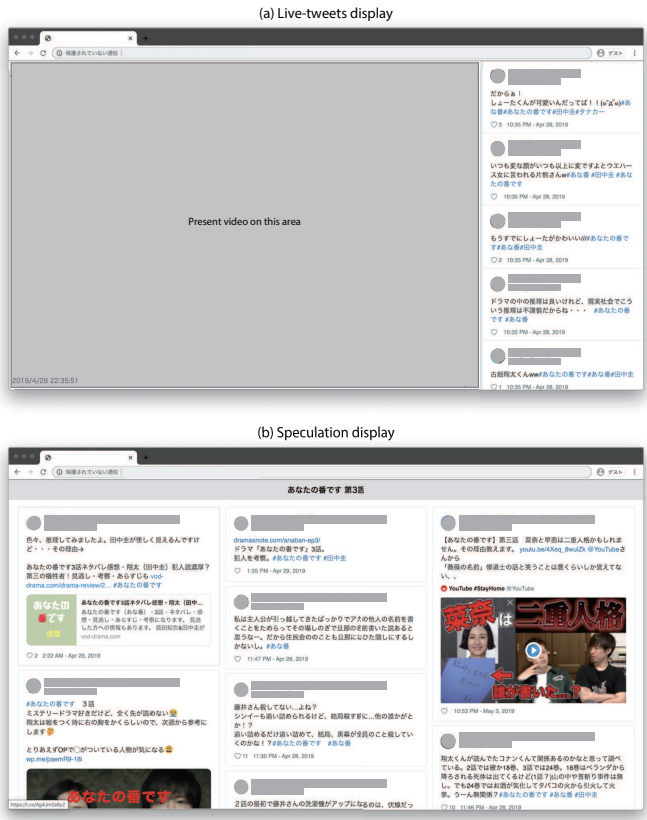}
    \end{center}
    \caption{Two prototypes presented to participants in our experiment (a: displays tweets in relation to a playback time within the episode; b: displays tweets containing speculative discussions after watching the episode).\protect\footnotemark}
    \label{fig:screenshots-prototypes}
\end{figure}

The interfaces of the two prototypes are presented in \figref{screenshots-prototypes}.
All tweets were displayed using the embedding function of Twitter so that participants can preview attached images or videos and follow hyperlinks in the tweets regarding both prototypes, though \textit{non-speculation-tweets} were less likely to contain such attachments according to our tweets analysis in \secref{tweets-analysis}.
Here, different from some previous proposals of second-screen applications, which we mentioned in \secref{interaction-techniques}, we designed these prototypes to be used in the main screen.
This is because some participants took part in this experiment remotely, as we mention later, and we could not guarantee that all of them can prepare similar second-screen devices.
Instead, all participants, including them, used these prototypes with a laptop.

\footnotetext{The usernames and icons are grayed out for anonymity.}

\subsection{Measures}
\label{sec:experiment-measures}

To evaluate the effect of the prototypes, we set up measures corresponding to our design strategies.
Our measures were guided by Fang et al. \cite{Fang2018}, who conducted a questionnaire-based user study to reveal how Danmaku interactions contributed to the enhancement of viewing experiences.

We adopted their questionnaire sets to measure social presence (based on \cite{Hassanein2007,Ou2014,Zhou2016}) and utilitarian value (based on \cite{Hsieh2008,Wang2013,Zhou2012}).
Because social presence is elicited by the feeling of being connected with others \cite{Fang2018,Zhou2016}, we expected that it can be enhanced not only by the live-tweets display but also by the speculation display, especially considering the findings described in \secref{interview-connectedness}.
On the other hand, utilitarian value refers to participants' perceived benefits from some systems \cite{Fang2018,Zhou2012}.
We thus expected that it is more related to the speculation display because, as we mentioned in \secref{interview-understanding}, the effect of deepening understandings of media content would be specific to speculation-related activities.

In addition, we followed Fang et al. \cite{Fang2018} to evaluating the quality of such experiences by measuring users' e-loyalty intention, i.e., their revisit intention and positive word-of-mouth referrals (based on \cite{Moon2013,Zhou2012}).
If our prototypes contribute to enhancing participants' consumption experiences, the e-loyalty intention would be increased, as they confirmed in the case of Danmaku interactions.
Here, the reason why we used this measure is that its prospective aspect has an affinity with speculation, as we mentioned that speculation is an ongoing experience based on a form of seriality in \secref{sharing-speculations}.
In other words, other measures that require retrospective evaluation, such as immersion \cite{DBLP:conf/tvx/RigbyBGC19}, would not fully reflect the effect of the speculation display and not optimal for discussing the pros and cons of the two prototypes.

In summary, from Fang et al.'s questionnaire \cite{Fang2018}, we adopted the same five items to measure social presence, four to measure utilitarian value, and four to measure e-loyalty intention, all of which were rated on a 7-point Likert scale ranging from 1 (strongly disagree) to 7 (strongly agree).
We note that these items were validated in their original studies, and thus, we analyzed the collected scores using the standard procedure involving ANOVA, as we describe later in \secref{experiment-result-quantitative}.

\subsection{Procedure}
\label{sec:experiment-procedure}

The experiment was conducted in a quiet room with a laptop we prepared, except for nine participants who took part in remotely.
For the remote participants, we sent URLs to the prototypes and instructed them to access the URLs using their laptops in accordance with the experimental steps.

Our experiment followed a within-participant design with three conditions: two prototypes and no intervention.
Participants watched the first through the third episode of \textit{Your Turn to Kill} (あなたの番です) in order with each condition.
The order of conditions was balanced across participants to avoid the influence of order effects or the content of each episode.
After participants finished watched each episode, they completed the questionnaire described above, except in the case of the speculation display (\figref{screenshots-prototypes}b).
In this case, we asked participants to freely explore the speculative discussions on the page to their satisfaction before filling out the questionnaire.
In addition, in the case of no intervention, we omitted the measure for utilitarian value in the same manner as Fang et al. \cite{Fang2018} because it was not applicable.
After participants watched the three episodes and filled out the questionnaire, we asked them for their comments about their consumption experiences and the overall experimental process, which lasted on average approximately 3.5 hours.

\subsection{Results}
\label{sec:experiment-result}

\subsubsection{Social Presence, Utilitarian Value, and E-loyalty Intention}
\label{sec:experiment-result-quantitative}

\begin{table}
    \caption{Result of a two-way ANOVA in participants' responses to measures for social presence.}
    \label{tbl:social-presence}
    \begin{minipage}{\columnwidth}
        \centering
        \begin{tabular}{lrrr}
            \toprule
            Effect                     & F      & df & $p$-value \\
            \midrule
            Episode                    &  0.623 & 2  & 0.541     \\
            Condition                  & 11.413 & 2  & $< 0.001$ \\
            Episode $\times$ Condition &  0.385 & 4  & 0.818     \\
            \bottomrule
        \end{tabular}
    \end{minipage}
\end{table}

\begin{table}
    \caption{Result of a two-way ANOVA in participants' responses to measures for utilitarian value.}
    \label{tbl:utilitarian-value}
    \begin{minipage}{\columnwidth}
        \centering
        \begin{tabular}{lrrr}
            \toprule
            Effect                     & F      & df & $p$-value \\
            \midrule
            Episode                    &  1.886 & 2  & 0.310     \\
            Condition                  & 18.778 & 1  & 0.001     \\
            Episode $\times$ Condition &  3.679 & 2  & 0.108     \\
            \bottomrule
        \end{tabular}
    \end{minipage}
\end{table}

\begin{figure}[tb]
    \begin{center}
        \begin{subfigure}{0.54\textwidth}
            \includegraphics[width=0.99\textwidth]{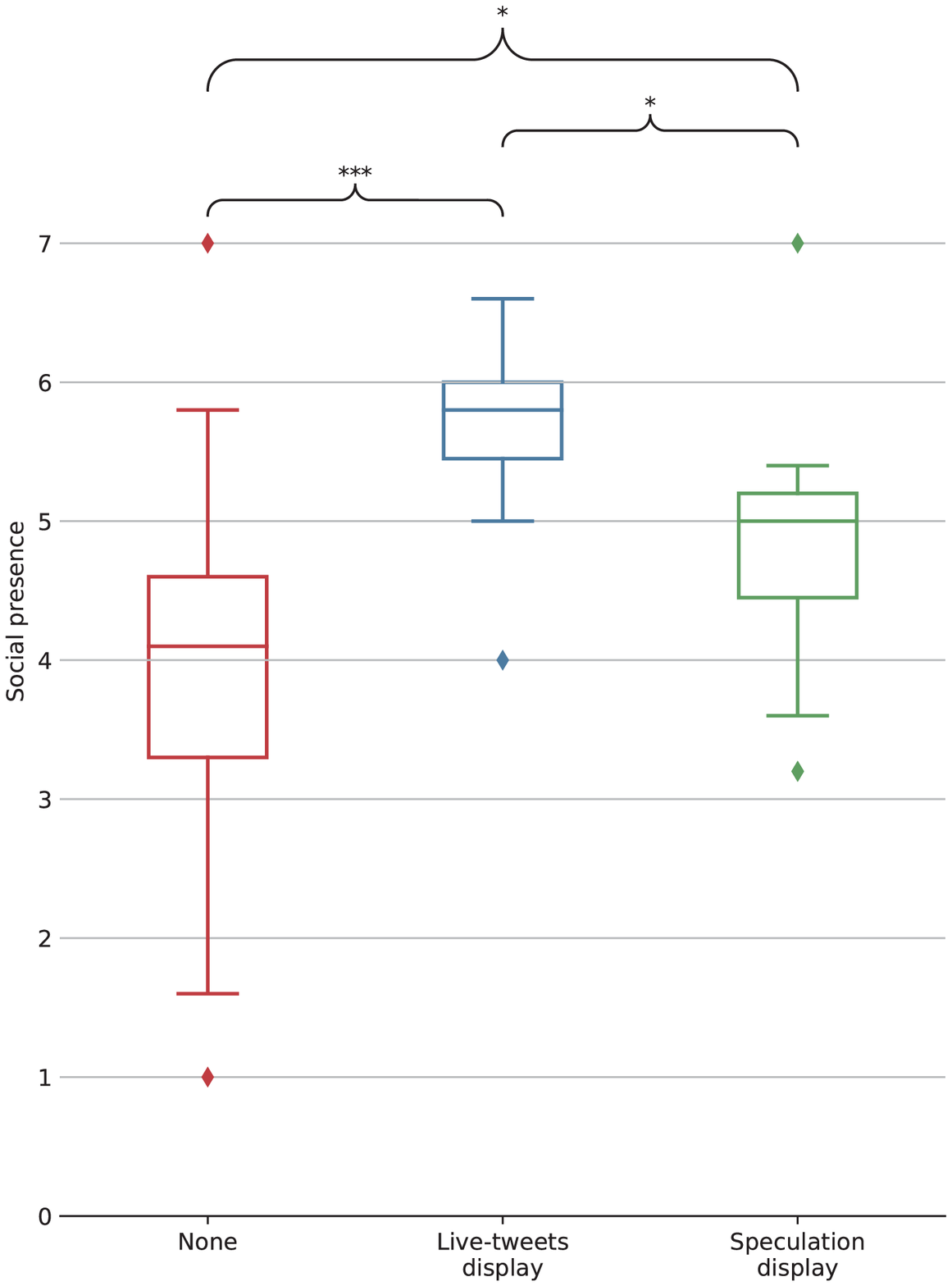}
        \end{subfigure}
        \begin{subfigure}{0.36\textwidth}
            \includegraphics[width=0.99\textwidth]{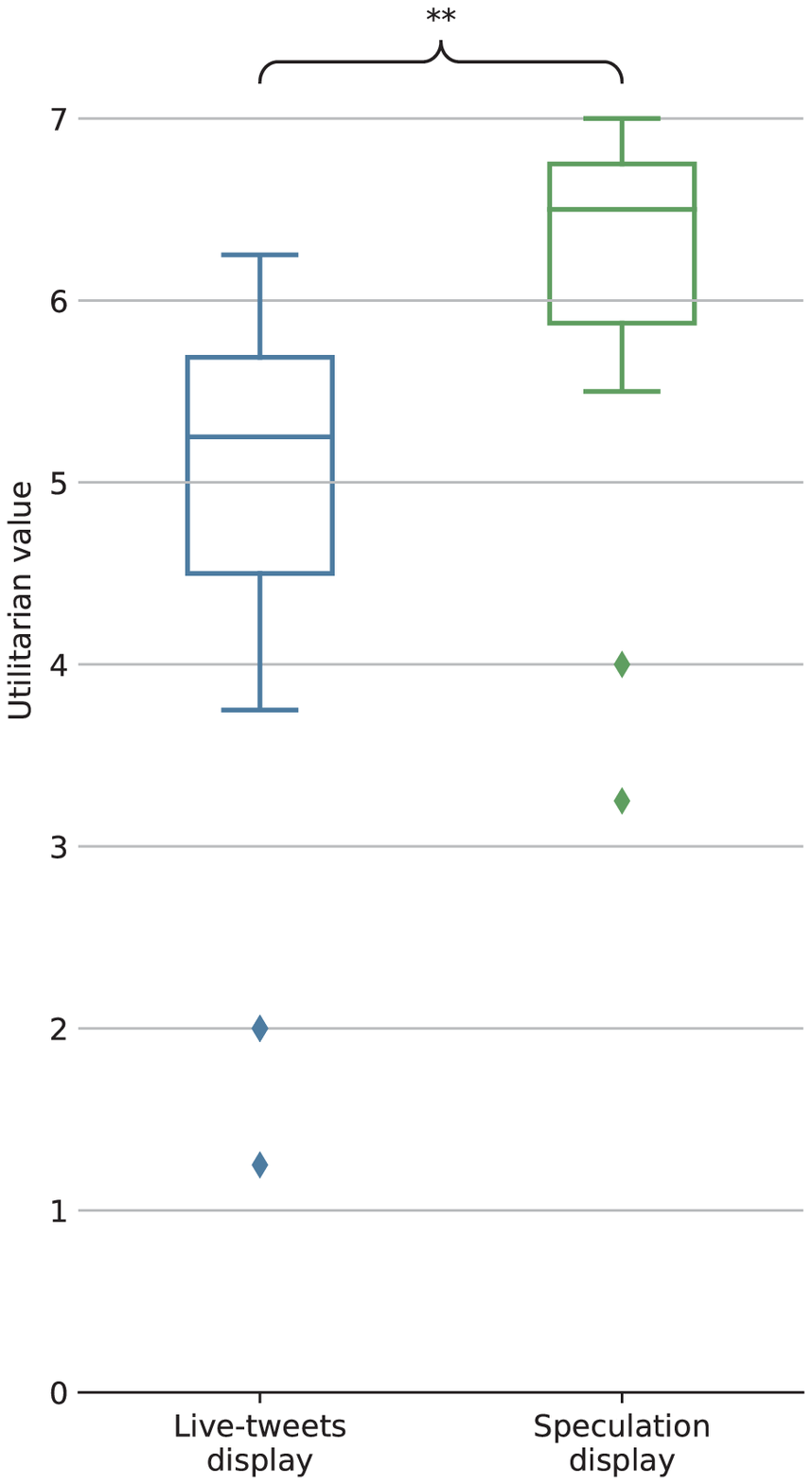}
        \end{subfigure}
    \end{center}
    \caption{Participants' responses to measures for social presence and utilitarian value.}
    \label{fig:social-utilitarian}
\end{figure}

For social presence, we confirmed that Levene's test did not found a significant difference in its variance ($p = 0.085$) and then conducted a two-way ANOVA.
As a result, a significant main effect of the conditions ($p < 0.001$) and an insignificant interaction effect ($p = 0.818$) were indicated, as presented in \tabref{social-presence}.
We thus conducted a post hoc test and there were significant effects between all pairs of conditions: no intervention and live-tweets display ($p < 0.001$); no intervention and speculation display ($p = 0.025$); and live-tweets display and speculation display ($p = 0.018$).
Similarly, for utilitarian value, we confirmed that Levene's test did not found a significant difference in its variance ($p = 0.091$) and then conducted a two-way ANOVA.
As a result, a significant main effect of the conditions ($p = 0.001$) and an insignificant interaction effect ($p = 0.108$) were indicated, as presented in \tabref{utilitarian-value}.
\figref{social-utilitarian} shows the scores indicating participants' sense of social presence and utilitarian value.

These results correspond with our expectations described in \secref{experiment-measures} and corroborate our findings presented in \secref{interview}.
First, the difference in social presence between the live-tweets display and no intervention conditions is consistent with the findings of the previous literature summarized in \secref{sharing-immediate-reactions}---namely, that engaging with immediate reactions during media consumption promotes the sense of being connected with others.
In particular, we confirmed our expectation that this effect was also present among catching-up users, which had not previously been studied in depth.

The two aspects of the positive effects of engaging with speculative discussions described in \secref{interview-understanding} and \secref{interview-connectedness} were also supported.
More specifically, the increase in utilitarian value, consistent with our expectation and participants' comments (discussed further in \secref{experiment-comments}), implies that the speculation display provided useful information and thus deepened participants' understandings of the content in comparison to the live-tweets display of immediate reactions.
In addition, the difference in social presence between the speculation display and no intervention conditions supports another effect: that speculative discussions can also provide a sense of connectedness with others.
Though this effect is not as powerful as with the live-tweets display, these points suggest that sharing speculations might offer enhanced consumption experiences to catching-up users.

\begin{table}
    \caption{Result of a two-way ANOVA in participants' responses to measures for e-loyalty intention.}
    \label{tab:social-presence}
    \begin{minipage}{\columnwidth}
        \centering
        \begin{tabular}{lrrr}
            \toprule
            Effect                     & F      & df & $p$-value \\
            \midrule
            Episode                    &  7.337 & 2  & 0.002     \\
            Condition                  &  1.463 & 2  & 0.242     \\
            Episode $\times$ Condition &  3.976 & 4  & 0.008     \\
            \bottomrule
        \end{tabular}
    \end{minipage}
\end{table}

\begin{figure}[tb]
    \begin{center}
        \includegraphics[width=0.72\textwidth]{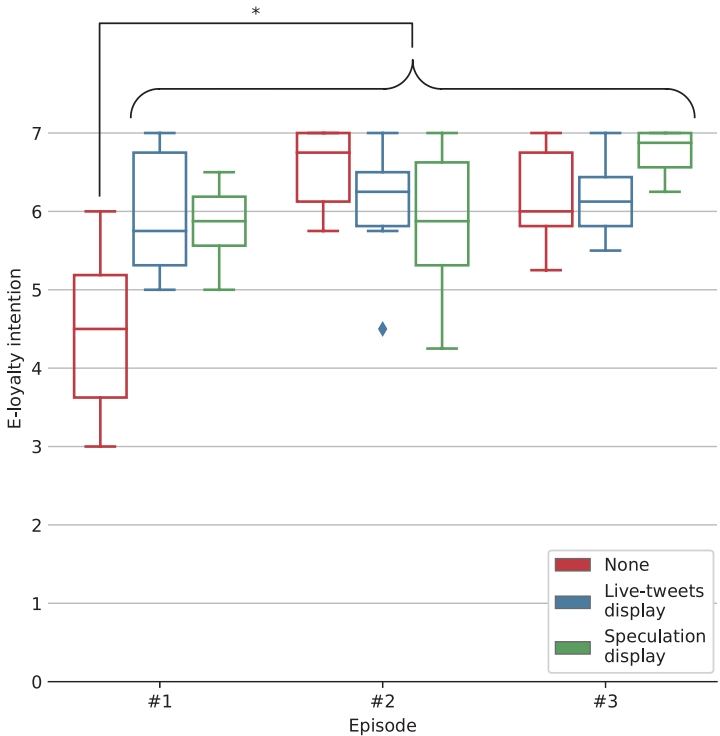}
    \end{center}
    \caption{Participants' responses to measures for e-loyalty intention.}
    \label{fig:e-loyalty}
\end{figure}

For e-loyalty intention, we confirmed that Levene's test did not found a significant difference in its variance ($p = 0.114$) and then conducted a two-way ANOVA.
As a result, a significant main effect of episode ($p = 0.002$) and a significant interaction effect ($p = 0.008$) were indicated.
We thus conducted a post hoc test across all nine combinations of the conditions and episodes, which indicated that the responses for watching the first episode without any intervention were significantly lower than all the other combinations.
\figref{e-loyalty} shows scores for participants' e-loyalty intention.

This result suggests that the participants' e-loyalty intention depends on the episode they watch. 
In particular, considering the distribution in \figref{e-loyalty}, the second and third episodes seemed to have attracted participants well regardless of the interventions so that participants would like to watch the next episode or spread positive word-of-mouth referrals.
On the other hand, in the first episode, the two prototypes effectively increased their e-loyalty intention, which implies that computational support can enhance the consumption experiences of catching-up users at least when the content has not fully attracted them yet.

\subsubsection{User Comments}
\label{sec:experiment-comments}

As shown above, our quantitative results demonstrated the distinctive effectiveness of our two prototypes.
In this section, we review the comments obtained from participants during the experiment to further explore our results.

All participants approved our prototypes and described their benefits, e.g., enhanced feelings of watching together with peers through the live-tweets display and a deepened understanding of the show's content through the speculation display:
\begin{quote}
    It was nice to be able to see how other people were watching at it and feel as if we were watching it together.
\end{quote}
\begin{quote}
    Watching the speculation for the next episode enabled me to appreciate the content from various perspectives. What's more, it helped me to deeply understand each of the many characters that appeared in the story.
\end{quote}
These comments align with the existing literature on live-tweets (\secref{sharing-immediate-reactions}) and our findings on speculation (\secref{interview-findings}).
Therefore, in combination with the results presented in \secref{experiment-result-quantitative}, we conclude that our prototypes enhanced participants' consumption experience, which suggests the potential of computational support for catching-up users.

In addition, some participants offered suggestions for improving the prototypes.
Although the feeling of being connected with others enabled by the live-tweets display was favored by many participants, five participants mentioned a desire to control the content presented on the display to some extent: 
\begin{quote}
    Sometimes I felt it a bit annoying that most of the tweets were about the coolness of the actor of the protagonist. I wanted to focus on the story by suppressing those tweets.
\end{quote}
\begin{quote}
    The live-tweets interface was not bad, but I would like to not see lots of reaction-like tweets such as ``oops'' and ``lol,'' but rather some meaningful tweets like ``I'm sure this person is lying because\ldots'' which can help me to think about the story more.
\end{quote}
These comments suggest a demand for the ability to customize the live-tweets display based on content. 
Such add-on functionality could be enabled by analyzing tweet data and applying clustering techniques.

Interestingly, two participants mentioned experiencing a dilemma between their desires to watch the next episode and to enjoy the speculation content:
\begin{quote}
    I really enjoyed the diverse speculations, some of which I had never imagined. However, the more articles I read, the more I was tempted to stop reading them and to watch the next episode to see which one would be correct.
\end{quote}
These comments are unique to catching-up users, who can binge watch an entire series \cite{Matrix_2014,FM6138} without exposing themselves to speculations after each episode.
Therefore, it is desirable to consider this unique predicament when we design applications for such users.
We discuss several possible options for addressing this point in the next section.

Another interesting comment was provided by one participant, who said:
\begin{quote}
    It is fun to go through so many different speculations. It would be even better if I could tell my opinion to others.
\end{quote}
It would be challenging to meet this demand in catch-up situations, as our prototypes cannot address it.
In other words, while our prototypes enhanced the consumption experiences of catching-up users, further technological developments would be desirable to close the gap in the experience of watching simultaneously with other users at the time of broadcast.

\section{Discussion}

\subsection{Collective Aspect of Speculation-based Media Consumption}

So far, we have revealed that engaging with online speculative discussions can contribute to deepening understandings of media content through both the semi-structured interviews (\secref{interview-understanding}) and the user experiment (\secref{experiment-result}).
In particular, the comments in the experiment (\secref{experiment-comments}) suggested that the participants obtained perspectives beyond the individual by appreciating speculations posted by various Internet users.
We would like to say that this structure can be aligned with \textit{collective intelligence} \cite{10.5555/550283}. 

In fact, Jenkins's early observation \cite{Jenkins2006} of the online speculative discussions of TV series pointed out the role of a fan community as collective intelligence.
We note that our observation was different from his discussion in terms of the lack of close-knit communities (e.g., a fan wiki), which can be attributed to the advent of social networking services.
However, the structure that people who relate to speculation exploit the Internet to search or join discussions about a sequel is common.
Also, in the same manner as Jenkins \cite{Jenkins2006} depicted, we observed that the well-considered speculations are shared by many users on the Internet and trigger further discussions on Twitter.

In this context, our prototype leveraging speculations can be understood to be served as an interface for harnessing collective intelligence.
That is, collecting related opinions from the Internet and displaying them is one of the basic approaches to leverage collective intelligence \cite{DBLP:journals/cacm/Gregg10}.
Given that, as many researchers have proposed sophisticated interfaces for harnessing collective intelligence \cite{DBLP:journals/cscw/GrassoC12,DBLP:conf/webi/VuMA13}, we anticipate that our prototype for supporting catching-up users could have been improved based on their designs.
Yet, as speculation is not only for deepening understandings of media content (\secref{interview-findings}), there is room for exploring how we can incorporate existing techniques for collective intelligence in terms of enhancing the consumption experiences of catching-up users based on speculation.

\subsection{Application Scenarios for Supporting Catching-Up Users}

Through the user experiment with our two prototypes, we confirmed that engaging with immediate reactions and speculative discussions are both beneficial to catching-up users, albeit with different effects.
This result suggests potential application scenarios for supporting catching-up users.
For example, catch-up TV services might leverage the combined effect of both approaches by offering users an interface to see live-tweets while they watch content and, after watching the episode, guiding them to explore speculative content before starting to watch the next episode.
In addition, sending a notification to a user who has stopped watching a series on an episode in the middle of a series with a link to a discussion article speculating about the consequent episodes can stimulate the user's desire to watch the next episode.

In addition to enabling catching-up users to see live-tweets and speculations posted by other users who watched the episode in sync with the broadcast, we can design and develop interactions that only catching-up users can appreciate.
Inspired by the recent proposals for story-based retrievals from TV shows \cite{DBLP:journals/ijmir/TapaswiBS15,DBLP:conf/iccv/ZhuKZSUTF15} and comics \cite{DBLP:journals/mta/MatsuiIAFOYA17,DBLP:conf/icdar/LeLBO15} or annotations \cite{DBLP:conf/www/DowmanTCP05,DBLP:journals/mta/PereiraSPSAP15}, we suggest that it may be helpful to guide catching-up users to be attentive by showing them a subtle alert during scenes about which others have frequently speculated.
This can be identified through an analysis of \textit{speculation-tweets}.
We acknowledge that such guidance can simply be prepared by the producer of the content without online speculative discussions.
However, as Gray and Mittell \cite{gray2007speculation} noted, speculation is often regarded as a ``game'' between viewers and producers.
As such, viewers may refuse spoilers that are officially released by producers.
In this way, such interactions could effectively enhance the consumption experiences of catching-up users without spoiling them.

In addition, this type of interaction could remedy the dilemma described in \secref{experiment-comments}.
Through leveraging the automatically retrieved relationships between scenes and speculations, it is possible to present speculations in the same manner as live-tweets.
Catching-up users can then enjoy both watching the content and speculating about it simultaneously and thus would not need to spend time to explore speculations between the intervals of the episodes.
Further studies are required to design and evaluate such interactions so that users can appreciate speculations without significant cognitive load while watching. 

\subsection{Avoiding Spoilers in Providing Speculation to Catching-Up Users}
\label{sec:disc-avoiding-spoilers}

Despite the promising applications of leveraging speculation to support catching-up users, our semi-structured interviews identified a potential problem: the risk of encountering spoilers (described in \secref{interview-spoilers}).
In this respect, many studies have examined spoiler detection and related interaction techniques \cite{DBLP:conf/coling/GuoR10,DBLP:conf/chi/Golbeck12,DBLP:conf/avi/NakamuraK12,DBLP:conf/asist/Boyd-GraberGZ13,DBLP:conf/cikm/YangJGZL19}.
For example, Guo et al. \cite{DBLP:conf/coling/GuoR10} proposed a method to detect spoiling movie reviews using latent Dirichlet allocation, while Golbeck \cite{DBLP:conf/chi/Golbeck12} demonstrated how simple keyword-based filtering using the name of actors or sports players could effectively block live-tweets containing spoilers, although the method's precision was poor.
Yang et al. \cite{DBLP:conf/cikm/YangJGZL19} proposed a spoiler detection method specifically designed for Danmaku comments that relied on a comment's similarity to comments made during the climax period, i.e., peak volume.

However, we argue that these methods are inadequate for supporting catching-up users.
As discussed in \secref{interview-findings}, the motivations behind this phenomenon of speculation are related both to social presence and to gain a deeper understanding of media content.
As such, any design that ignores false positives (i.e., that filters non-spoiler information) would not desirable, as it would run the risk of missing a large amount of information important for deepening the understanding.
In addition, a criterion for filtering spoilers should be made dynamic: Depending on which episode the user has finished watching, the criteria should change so that as much speculation as possible is provided about the story prior to the episode, while eliminating any information to be revealed in later episodes.

We note that Jones et al. \cite{DBLP:journals/jodl/JonesNS18} proposed a method to avoid spoilers on a fan wiki that allowed a user to specify an episode and regarded information disclosed before that episode as known by the user in advance.
However, this method relies on the Memento extension \cite{RFC_7089} for the HTTP protocol, which is not yet globally available.
In addition, fan wikis are not always available for various media.

Therefore, we conclude that further developments to address the problem of spoilers are required to offer better consumption experiences to catching-up users.
For example, if we can assume that our data have reliable timestamps, the simple approach of filtering based on airtime would suffice, as in \secref{experiment-prototypes}.
Alternatively, it may be possible to determine topics that appeared around the airtime of an episode by employing methods for detecting emerging topics \cite{Kontostathis2004,10.1145/1814245.1814249}.
Filtering such topics for catching-up users who have not yet watched the episode would help them avoid spoilers.

\subsection{Limitations and Future Work}

Although our results shed light on the unique characteristics of speculation in contemporary media consumption and confirmed the positive effects of our prototypes for catching-up users, the participants in this study were mainly from East Asian cultures.
In addition, the fact that our experiment involved participants who took part remotely, as mentioned in \secref{experiment-procedure}, would raise some concern on the ecological validity, as it eliminated the use of second-screen devices.
To generalize our results, further investigations involving participants in other cultures and considering various use cases are desirable.

At the same time, seeking ways to support speculations for diverse media content, not only TV series from outside of East Asian countries but also other media types, would be interesting.
For example, considering the importance of books as a form of media, future studies might develop a prototype to enhance readers' experiences of catching up on series of comics or novels by designing how catching-up readers engage with speculations.

Another limitation is related to data collection---namely, that our proposed two prototypes relied on the assumption that there were enough live-tweets and speculations on Twitter to entertain users.
This assumption may not always hold true, given that there is a wide range of media content, including ones not so popular.
However, especially for speculation, its source is not limited to Twitter but also includes other social networks, such as YouTube, blogs, and Instagram, as observed in \secref{interview-findings} and \secref{tweets-analysis}.
This would allow us to gather speculations to be presented in other ways.
Therefore, future work is demanded to establish such an approach while taking into account the spoiler problem discussed in \secref{disc-avoiding-spoilers}.

\section{Conclusion}

In this paper, we focused on the increase in the number of catching-up users and the emergence of a new media consumption experience centering on speculation, elucidating new approaches to enhancing the experiences of the catching-up users.
We first conducted semi-structured interviews to figure out how people are engaging with speculation during media consumption, which revealed that two positive effects (i.e., deepening understanding and sense of connectedness with others) as well as the possible risk of encountering spoilers.
We then conducted a quantitative analysis using public tweet data to provide background for discussing computational supports for catching-up users, which illustrated the unique aspects of \textit{speculation-tweets}, including their timing, length, and content.
Finally, we performed a user experiment to evaluate the effect of two different approaches based on immediate reactions and speculations, which demonstrated their effectiveness in enhancing consumption experiences.
Our results and discussions lay the groundwork for providing catching-up users with computational support, particularly by leveraging speculation, which has not been well explored to date despite its effectiveness.

%
\begin{acks}
This work was partially supported by JST ACT-X, Grant Number JPMJAX200R, Japan.
\end{acks}

%
\bibliographystyle{ACM-Reference-Format}
\bibliography{reference}

\received{June 2020}
\received[revised]{October 2020}
\received[accepted]{December 2020}

%
\end{document}